\documentclass[fleqn,usenatbib]{mnras}

\usepackage{newtxtext,newtxmath}
\usepackage{natbib} 

\usepackage[T1]{fontenc}

\DeclareRobustCommand{\VAN}[3]{#2}
\let\VANthebibliography\thebibliography
\def\thebibliography{\DeclareRobustCommand{\VAN}[3]{##3}\VANthebibliography}

\usepackage{graphicx}	
\usepackage{amsmath}	

\title[The first optical study of IGR J17091$-$3624]{Exotic optical variability in the black hole X-ray binary IGR J17091$-$3624}

\author[Saikia et al. 2025]{Payaswini Saikia$^{1,2}$\thanks{E-mail: payaswini.saikia@yale.edu}, 
David M. Russell$^{2}$, D. M. Bramich$^{2}$, Kevin Alabarta$^{2}$, Sandeep Rout$^{2}$, Federico \newauthor Vincentelli$^{3,4}$, Mariano Mendez$^{5}$, Diego Altamirano$^{6}$, Federico Garcia$^{7}$, 
M. C. Baglio$^{8}$, Fraser \newauthor Lewis$^{9,10}$ and Yi-Jung Yang$^{11,12}$ \\
\\
$^{1}$Department of Astronomy, Yale University, PO Box 208101, New Haven, CT 06520-8101, USA\\
$^{2}$Center for Astrophysics and Space Science (CASS), New York University Abu Dhabi, PO Box 129188, Abu Dhabi, UAE\\
$^{3}$Instituto de Astrofísica de Canarias, E-38205 La Laguna, Tenerife, Spain\\
$^{4}$Departamento de Astrofisica, Universidad de La Laguna, 38206 Santa Cruz de Tenerife, Spain\\
$^{5}$Kapteyn Astronomical Institute, University of Groningen, PO Box 800, NL-9700 AV Groningen, the Netherlands\\
$^{6}$School of Physics and Astronomy, University of Southampton, Southampton, SO17 1BJ, UK\\
$^{7}$Instituto Argentino de Radioastronomía (CCT La Plata, CONICET; CICPBA; UNLP), C.C.5, (1894) Villa Elisa, Buenos Aires, Argentina\\
$^{8}$INAF, Osservatorio Astronomico di Brera, Via E. Bianchi 46, 23807 Merate (LC), Italy\\
$^{9}$Faulkes Telescope Project, School of Physics and Astronomy, Cardiff University, The Parade, Cardiff, CF24 3AA, Wales, UK\\
$^{10}$The Schools' Observatory, Astrophysics Research Institute, Liverpool John Moores University, 146 Brownlow Hill, Liverpool L3 5RF, UK\\
$^{11}$Graduate Institute of Astronomy, National Central University, 300 Zhongda Road, Zhongli, Taoyuan 32001, Taiwan\\
$^{12}$Laboratory for Space Research, The University of Hong Kong, Cyberport 4, Hong Kong
}

\date{Accepted 2026 January 18. Received 2025 December 17.}

\pubyear{2026}

\begin{document}
\label{firstpage}
\pagerange{\pageref{firstpage}--\pageref{lastpage}}
\maketitle

\begin{abstract}

IGR J17091–3624 is a distinctive black hole X-ray binary exhibiting exotic variability, including complex “heartbeat” oscillations in its X-ray light curves, similar to those observed in GRS 1915+105, a system renowned for its structured, rapid X-ray variability but heavily obscured at optical wavelengths. In contrast, IGR J17091-3624 is less obscured, making it a more accessible target for optical investigations. Due to its weak radio emission, optical and infrared data are essential to probe the jet and outer disc behavior of IGR J17091-3624. This study presents the first long-term optical monitoring of IGR J17091-3624, using data from the Las Cumbres Observatory (LCO) over its 2011, 2016, and 2022 outbursts. We combine these observations with quasi-simultaneous X-ray data from Swift/XRT, RXTE, and NICER, employing light curve and variability analysis, spectral energy distributions, color-magnitude diagrams, and optical/X-ray correlations to investigate optical emission mechanisms. We find that the optical and X-ray fluxes are significantly correlated, following a power-law relation ($F_{\mbox{\scriptsize opt}} \propto F_{\rm X}^{0.40\pm0.04}$), suggesting that the optical emission in IGR J17091-3624 is dominated by an X-ray-irradiated accretion disk. Based on optical spectral slope constraints, we estimate the extinction toward IGR J17091-3624 as $A_{V} =$ 4.3 to 6.6 mag, which translates to $N_{\mbox{\scriptsize H}}$ = 1.3--1.9 $\times 10^{22}\, \rm cm^{-2}$. The global optical/X-ray correlation suggests a distance estimate of 8--17 kpc, in line with previous findings. High-cadence optical observations show tentative evidence of optical oscillations that may arise from reprocessed X-ray modulations, although confirming this will require higher time-resolution optical data.

\end{abstract}

\begin{keywords}
accretion, accretion disks --- black hole physics --- ISM: jets and outflows --- X-rays: binaries – X-rays: individual: IGR J17091-3624
\end{keywords}


\section{Introduction} \label{sec:intro}

Black hole X-ray binaries (BHXBs) are binary systems comprising a stellar-mass black hole (BH) that accretes material from a secondary donor star through an accretion disk enveloping the central BH. The majority of BHXBs are transient in nature, predominantly dwelling in quiescence with minimal accretion. However, they sometimes enter outbursts characterized by a substantial increase in accretion rate and luminosity, brightening by orders of magnitude at all wavelengths. During an outburst, distinct spectral states emerge, notably the hard state (HS), the soft state (SS), and the intermediate states \citep[see e.g.][]{mcclintock2003black,Belloni2010}. In the HS, the X-ray spectrum is dominated by a power-law component. In the SS, thermal emission from an accretion disk becomes the primary contributor to the energy spectrum, characterized by variability with low fractional rms. The intermediate states occur during the transition between HS and SS, representing a mix of both spectral components. The HS and hard-intermediate state (HIMS) exhibit robust broadband noise and type-C quasi-periodic oscillations \citep[QPOs,][]{ingram} in the power density spectra (PDS). The soft-intermediate state (SIMS) typically displays red noise and type-B QPOs \citep{abc}. The spectral evolution of BHXBs often traces a `q-shaped' hysteresis loop in the hardness--intensity diagram \citep[HID,][]{my,homanhid,f2009,b2010}, moving counterclockwise \citep[although see][for rare exceptions]{dunn2010,saikia1910mnras}.

Outbursts are often accompanied by the ejection of compact, steady jets in the HS and ballistic jets during the transition to the SS \citep{f2009}. A robust observational link between accretion and ejection during the HS is evident, demonstrated by the correlation between the radio and X-ray emission of BHXBs \citep[e.g.][]{gallofenderpooley2003,corbel2003,Corbel2013,esp,gallo2008rx}, a relationship that extends to supermassive BHs \citep[e.g.][]{merloni2003,Falcke2004,saikia2015,saikia2018}. A similar global correlation is also observed between the optical/infrared and X-ray emission of BHXBs in the HS, roughly following the relation $L_{\rm OIR} \propto L_{\rm X}^{0.6}$ for an X-ray irradiated accretion disk, spanning eight orders of magnitude in X-ray luminosity \citep{Russell2006,Russell2007,Coriat2009,Shahbaz2015}.

The BHXBs GRS~1915+105 and IGR~J17091-3624 are two remarkable systems known for exhibiting high-amplitude, highly structured sub-second timescale variability in the X-ray energy bands \cite[see e.g.][]{diego2011,wang2,wang1}. 
The variability is a characteristic predicted by the thermal-viscous instability of the accretion disk and can be theoretically described by accretion disk instability driven by dominant radiation pressure \cite[see e.g.][]{l1,belloni97,l2,l3,l4}. This is likely the result of recurring episodes of rapid matter depletion and replenishment in the inner disk, possibly accompanied by ejections during the recovery from X-ray dips \citep{belloni97}. Recent studies indicate that at high accretion rates, disk instabilities can drive relativistic ejections during state transitions, leading to continuous disk depletion and replenishment \citep{vinnature}. These X-ray variations make IGR~J17091-3624 and GRS 1915+105 stand out as two uniquely peculiar BHXBs. However, while both have been extensively investigated in the X-rays over different outbursts \citep[see e.g.][]{diego2011,rodri,king2012,db,rao, iyer,yanan,wang2,wang1}, there are not many detailed optical/UV/infrared studies of these sources.

While GRS 1915+105 exhibits a rich range of X-ray variability, none have been observed at optical wavelengths, as the extreme extinction towards the source ($A_V = 19.6 \pm 1.7$ mag; \citealt{cha}) prevents detection of any optical variability. Radio observations of GRS 1915+105 throughout different outbursts show the presence of a self-absorbed compact jet in the HS, and discrete ejections associated with the state transition, with detections at the 1–2 mJy level \citep{rodri}. Radio emission was not detected in the high/soft state, with an upper limit corresponding to the image background rms of $\sim$ 0.13 mJy \citep{king2012}. \cite{neil} monitored GRS 1915+105 for 7 years in the infrared (K-band) and reported positive correlations between infrared flux and X-ray flux. GRS 1915+105 showed slow infrared oscillations, with the infrared emission highly correlated with radio flares, hinting at the origin of such flares being related to synchrotron emission coming from ejections \citep{fen97}. GRS 1915+105 provided the first clear evidence of jet emission in the infrared in a BHXB. Infrared flares in GRS 1915+105 were found to be associated with dips in X-rays \citep{eik98}. A detailed study of the evolution of the X-ray and infrared cross-correlation function (CCF) in this source reported consistently significant anti-correlations during the high X-ray variability epochs, with the X-ray preceding the infrared by $\sim$13$\pm$2 seconds \citep[][]{lasso}. Recent studies have reported both long- and short-term mid-infrared variability in GRS 1915+105 using WISE and JWST data, respectively, with fractional rms amplitudes of $\sim$1 per cent \citep{2025MNRAS.537.1385G}.

\subsection{IGR J17091$-$3624}  \label{sec:intro2}

The transient microquasar IGR J17091$-$3624 (hereafter IGR J17091) is a distinctive BHXB that exhibits peculiar X-ray variability patterns, with relatively low extinction along its line of sight compared to GRS~1915+105, making it better suited for optical studies. It was discovered in 2003 with the INTErnational Gamma-Ray Astrophysics Laboratory (INTEGRAL) while in outburst \citep[][]{dis1}. Archival data from different X-ray missions confirmed that the source had previous outbursts in 1994, 1996 and 2001 \citep[][]{dis2,dis3,dis4}, while new outbursts were later detected in 2007 \citep{dis2007}, 2011 \citep[][]{dis5,dis6} and 2016 \citep[][]{dis7}. Each of these outbursts lasted a few months to a few years, with a quiescent period typically spanning four years between these episodes. X-ray monitoring with the Neil Gehrels \textit{Swift}/BAT, and follow-up observations with \textit{NICER} detected another outburst of IGR~J17091 in 2022 March \citep[][]{dis2022}. The source entered a new outburst in February 2025, first detected by INTEGRAL \citep{integral}, and subsequently confirmed by NICER \citep{nicer} and optical observations with LCO \citep{katerina}. The outburst likely ended in the optical bands around the end of April 2025 \citep{newopt}, although faint X-ray activity persists till June 2025 \citep{newx}.

The lack of a detailed study of the optical counterpart in quiescence makes it difficult to determine key parameters of IGR J17091, such as the BH and donor star masses, orbital inclination, and distance. However, approximate estimates have been obtained through alternative methods. The inclination angle of the system is between 50$^{\circ}$ to 70$^{\circ}$ with the upper limit determined by the absence of any eclipses \citep[][]{king2012}. It has a low-spin or retrograde BH \citep{rao, yanan}. The parallax distance to the source has not been measured. IGR~J17091 is estimated to be within the range of 11-17 kpc from the luminosity at the HS to SS transition, assuming a BH mass of 10 $M_{\odot}$ \citep{rodri,iyer}. However, the mass of the compact object as well as the companion star in IGR~J17091 are still a matter of debate. Given the similarities between IGR~J17091 and GRS 1915+105, one can assume that the source accretes at a high-Eddington rate. This implies that IGR~J17091 probably harbors a very small BH with mass $\sim 3M_{\odot}$, unless it either has a neutron star (NS) primary, or is very distant with $d >$ 17 kpc \citep[][]{diego2011}. Various papers suggest a much higher mass in the range of 8.7-12.3 $M_{\odot}$ \citep[see e.g.][]{db,iyer}, thus strengthening the argument that it probably has a large distance with $d >$ 17 kpc.

For IGR J17091, \cite{chaty2008} initially reported the discovery of two blended candidate infrared counterparts C1 and C2, separated by $\sim$1~arcsec, favouring C1 as the more likely counterpart. This finding was later revised by \cite{Torres2011} with optical/infrared observations from the 6.5-m Magellan Baade telescope (Las Campanas Observatory). They compared optical imaging of IGR~J17091, acquired during both quiescence and outburst, to successfully identify the optical counterpart at R.A. 17:09:07.62 and Dec. $-$36:24:25.35 (J2000). They then used infrared imaging acquired during quiescence along with the coordinates of the optical counterpart to establish C2 as the infrared counterpart. Furthermore, \cite{Torres2011} found that C2 is actually a blend of two unresolved point-like sources of similar brightness, separated by 0.4~arcsec, with the slightly brighter source identified as the true infrared counterpart (with coordinates that match those of the optical counterpart). Due to the presence of contaminating sources in the close vicinity of IGR~J17091, any optical/infrared analysis must take care to account for blending, especially at, or close to, quiescence.\\

Despite the challenges, a detailed optical study of IGR~J17091 is important because until now it is the only known "heart-beating" source that can be studied in the optical, as the extinction towards GRS 1915+105 is extremely high \citep[][]{cha}. Understanding the optical behavior of these sources is key to revealing the full physics of accretion and the properties of accretion disks in BHXBs. Their high variability and unusual behavior challenge current accretion theories, offering insights that may uncover new physics in BH accretion and jet dynamics.

In this paper, we present the first long-term optical monitoring of IGR~J17091 using the Las Cumbres Observatory (LCO) network, covering its three recent outbursts in 2011, 2016, and 2022. To investigate the multi-wavelength evolution of the source, we analyze quasi-simultaneous X-ray observations from Swift/XRT, RXTE, and NICER. Furthermore, we report new mid-infrared observations obtained with VISIR during the 2022 outburst and incorporate near-infrared detections from the literature to construct a broadband spectral energy distribution (SED), providing further insights into the emission mechanisms at play. The observations and data used in this study are detailed in Section \ref{sec:obs}. In Section \ref{sec:results}, we present our findings, including the source evolution, spectral energy distribution, color-magnitude diagram, and optical/X-ray correlations. Section \ref{sec:dis} provides an interpretation and discussion of these results, while Section \ref{sec:con} summarizes our conclusions.

\begin{figure*}
    \centering
    \includegraphics[width = 2\columnwidth]{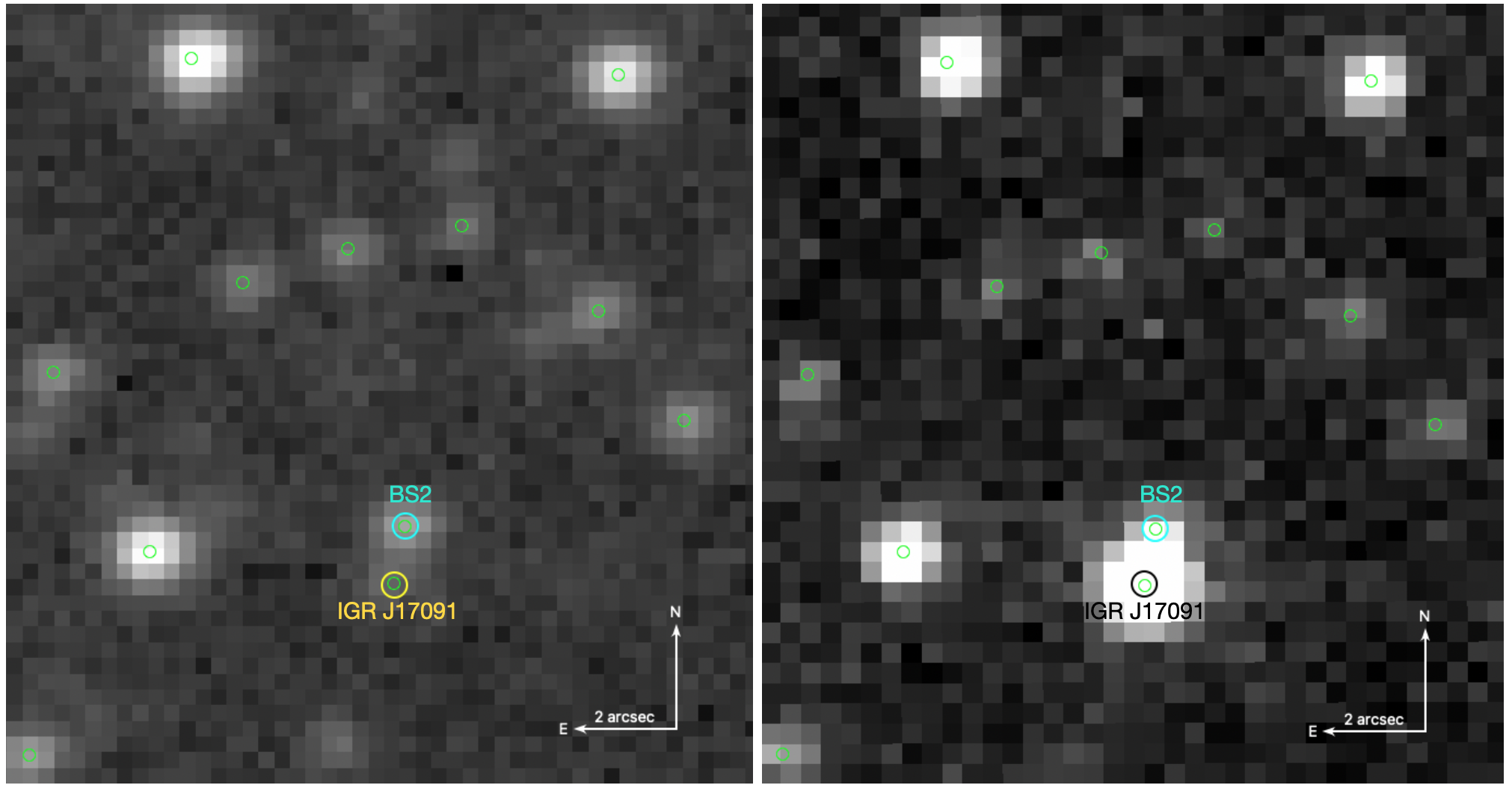}
    \caption{Small cut-outs from LCO $r^{\prime}$-band images of the field of IGR~J17091, acquired on different nights under good seeing conditions. \textit{Left:} Image acquired at SSO with the 2-m Faulkes telescope on 2021 April 19 (MJD 59323). The seeing is 0.95$^{\prime\prime}$. IGR~J17091 is in quiescence and marked with a yellow circle (0.5$^{\prime\prime}$ diameter). \textit{Right:} Image acquired at CTIO with one of the 1-m telescopes on 2022 May 27 (MJD 59726). The seeing is 1.23$^{\prime\prime}$. IGR~J17091 is in outburst and marked with a black circle (0.5$^{\prime\prime}$ diameter). \textit{Both:} The nearest resolved source to IGR~J17091 is marked with a cyan circle (BS2). The green circles indicate \textit{Gaia}~DR3 sources in the field.}
    \label{fig:oqfc}
\end{figure*}

\section{Observations} \label{sec:obs}

\subsection{Optical monitoring} \label{sec:obsLCO}

IGR~J17091 has been monitored at optical wavelengths with the Las Cumbres Observatory \citep[LCO;][]{Brown2013} since 2011, as part of an on-going monitoring campaign of $\sim$50 LMXBs co-ordinated by the Faulkes Telescope Project \citep{Lewis2008}. Imaging was carried out primarily in the SDSS $i^{\prime}$, $r^{\prime}$ and Bessell $V$ filters with the 2-m Faulkes Telescopes at the Haleakala Observatory (Maui, Hawai`i, USA) and the Siding Spring Observatory (SSO, Australia). Further imaging was obtained with the 1-m telescopes at the Siding Spring Observatory, the Cerro Tololo Inter-American Observatory (CTIO, Chile), and the South African Astronomical Observatory (SAAO, South Africa). 

The optical imaging data for the target IGR~J17091 were reduced by the LCO data reduction pipelines ORAC and BANZAI \citep{banzai}, for data from before and after April 2016, respectively. Following this, they were ingested into the LCO archive.\footnote{\url{https://archive.lco.global}} In near real-time, our data analysis pipeline ``X-ray Binary New Early Warning System (XB-NEWS)'' \citep{Russell2019,Goodwin2020} downloads all of the available reduced science images of the target (including related calibration files) from the LCO archive and performs quality control on the images to ensure that only images of good enough quality are processed further. Our cut-off date for downloaded imaging data for IGR~J17091 is 2024 Oct 27 (MJD 60610). In addition to the science images rejected automatically by the XB-NEWS quality control, we manually rejected six images in the $r^{\prime}$-band that were unusable due to various issues such as too many bad pixels, poor flat-field correction, and other artefacts. 

\subsubsection{Data processing} \label{sec:dataproc}

The XB-NEWS pipeline processes each reduced science image independently. For a specific image, it starts by detecting the sources in the image using SExtractor \citep{sx}, employing a detection threshold set to the estimated sky background plus 2.5$\times$ the estimated sky noise.\footnote{SExtractor requires five connected pixels in a ``plus'' shape to all exceed this threshold for a source detection to be declared.} The detected sources are used to compute a median value of the point-spread function (PSF) full-width at half-maximum (FWHM), which is taken as an estimate of the image PSF FWHM. They are also matched against Gaia DR2\footnote{\url{https://www.cosmos.esa.int/web/gaia/dr2}} source positions using {\tt astrometry.net} \citep{net}, and an astrometric solution for the image is derived from the matches. From the astrometric solution, XB-NEWS determines the nearest source detection to the known target coordinates, taken to be R.A. 17:09:07.605 and Dec. $-$36:24:25.35 (ICRS; reference epoch J2015.5) for IGR~J17091 (see Sec.~\ref{sec:dan}). If this source detection is within 1$^{\prime\prime}$ of the target coordinates, then XB-NEWS adopts it as the target detection for the image. Otherwise, a repeat run of SExtractor is performed on the image at a lower detection threshold than previously (the estimated sky background plus 1.0$\times$ the estimated sky noise), and a new nearest source detection to the target coordinates is determined. Again, if this source detection is within 1$^{\prime\prime}$ of the target coordinates, then XB-NEWS adopts it as the target detection for the image, otherwise an entry for the target is appended to the image source list using the target coordinates (to enable subsequent forced photometry). In summary, the final image source list consists of all sources detected in the reduced science image at the original detection threshold, and a target entry that is either a source detection (at one of two possible thresholds) or a nominal set of target coordinates.

XB-NEWS then performs photometry on the reduced science image for all of the sources in the image source list. Specifically, it performs multi-aperture photometry \citep{Stetson1990}, and standard (fixed) aperture photometry for each of the aperture radii from the set 0.5$\times$, 0.75$\times$, 1$\times$, 1.5$\times$, 2$\times$, and 2.5$\times$ the image PSF FWHM. The purpose of defining the aperture radii for the aperture photometry as multiples of the image PSF FWHM is to maintain approximately constant aperture losses for each source across all images. The apertures for both types of photometry are centred on the estimated coordinates of the sources in the image source list. Note that if the target entry in the source list is not a formal detection above the detection threshold, then the target photometry on the image is forced photometry at the target coordinates.

With the astrometry and photometry completed for all of the images from a particular observing configuration (i.e. those from the same site, telescope, instrument, filter, binning, and LCO pipeline), XB-NEWS constructs the (uncalibrated) lightcurves for all of the detected sources from these images (unnecessary for the target, since it has already been identified in all images). This is done using the {\tt DBSCAN} spatial clustering algorithm \citep{ester}. Flux calibration of the lightcurves is then achieved as follows. First, sources with lightcurves that have a number of epochs that is at least one third of the number of images are selected and matched with the standard stars in our own extended version of the ATLAS-REFCAT2 photometric catalogue \citep{Tonry2018}, which includes other catalogues such as Pan-STARRS1 DR1 \citep{pan} and APASS DR10 \citep{apass}. Then, XB-NEWS fits a photometric model to the lightcurves (in magnitudes) of these selected sources (both the matched and unmatched ones), that includes spatially variable zero-point offsets for each image, polynomial terms that depend on source PSF FWHM and ellipticity, and mean magnitudes as free parameters for those sources without a match in our standard star catalogue \citep{Bramich2012}. The fit is iterated twice with outlier down-weighting to guard against variable sources from degrading the solution. Note that XB-NEWS does not include any colour terms in the photometric model since close-in-time multi-band observations are not guaranteed to be available to the pipeline in general. However, given the close match between the passbands of the observational and standard systems, the absence of colour terms will only introduce small systematic errors ($<1-2\,$\%) in the absolute calibration of the photometry for most sources, and have no adverse effect on the relative calibration of the photometry for a single source (i.e. on the shape of each lightcurve). For the SDSS $i^{\prime}$ and $r^{\prime}$-band photometry, XB-NEWS employs the Pan-STARRS1 $i_{\mbox{\scriptsize P1}}$ and $r_{\mbox{\scriptsize P1}}$ standard magnitudes,\footnote{As listed in ATLAS-REFCAT2, having been computed as a weighted average of standard magnitudes from various other catalogues. In our case, Pan-STARRS1 DR1 does not cover the field of IGR~J17091, and consequently these standard magnitudes do not include Pan-STARRS1 DR1 standard magnitudes in their computation.} respectively, from our catalogue in the fit, while for the Bessell $V$-band photometry, the Johnson $V$ standard magnitudes are employed in the fit. Note that the Pan-STARRS1 $i_{\mbox{\scriptsize P1}}$ and $r_{\mbox{\scriptsize P1}}$ standard magnitudes are on the AB system, while the Johnson $V$ standard magnitudes are on the Vega system. As the final step in the flux calibration, XB-NEWS uses the fitted photometric model to calibrate all of the lightcurves, including that of the target, for the particular observing configuration under consideration.

In advance of the deblending analysis in Sec.~\ref{sec:dan}, it is necessary to mention some final details about the fluxes, magnitudes, and zero-point offsets determined by XB-NEWS. Firstly, XB-NEWS employs the following equation with arbitrary zero-point to convert between the instrumental flux $f_{\mbox{\scriptsize ins}}$ (ADU/s) and instrumental magnitude $m_{\mbox{\scriptsize ins}}$ measured for a source in an image:
\begin{equation}
m_{\mbox{\scriptsize ins}} = 25 - 2.5 \log_{\mbox{\scriptsize 10}} \left( f_{\mbox{\scriptsize ins}} \right)
\end{equation}
Secondly, the specific zero-point offset $\Delta z$ from the fitted photometric calibration model corresponding to the source brightness measurement is used to calibrate $m_{\mbox{\scriptsize ins}}$ to a standard magnitude $m_{\mbox{\scriptsize std}}$ as follows:
\begin{equation}
m_{\mbox{\scriptsize std}} = m_{\mbox{\scriptsize ins}} - \Delta z
\end{equation}
Now consider the same source at the top of the atmosphere with flux $f_{\mbox{\scriptsize std}}$. Then:
\begin{equation}
m_{\mbox{\scriptsize std}} = z_{\mbox{\scriptsize std}} - 2.5 \log_{\mbox{\scriptsize 10}} \left( f_{\mbox{\scriptsize std}} \right)
\end{equation}
where $z_{\mbox{\scriptsize std}}$ is the zero-point of the standard photometric system (and whose value depends on the units adopted for $f_{\mbox{\scriptsize std}}$). We collect all of the flux loss for the source along the light path from the top of the atmosphere through to the creation of the image, including the flux unit conversion to ADU/s, into a single factor $\kappa_{\mbox{\scriptsize sys}}$. We also use $\omega_{\mbox{\scriptsize apr}}$ to represent the fraction of the source flux in the image that falls within the (synthetic) aperture used for the XB-NEWS aperture photometry. Then we may relate $f_{\mbox{\scriptsize ins}}$ and $f_{\mbox{\scriptsize std}}$ via:
\begin{equation}
f_{\mbox{\scriptsize ins}} = \omega_{\mbox{\scriptsize apr}} \, \kappa_{\mbox{\scriptsize sys}} \, f_{\mbox{\scriptsize std}}
\end{equation}
Doing some algebra with equations~(1)$-$(4), then one may derive:
\begin{equation}
\Delta z = 25 - z_{\mbox{\scriptsize std}} - 2.5 \log_{\mbox{\scriptsize 10}} \left( \omega_{\mbox{\scriptsize apr}} \right) - 2.5 \log_{\mbox{\scriptsize 10}} \left( \kappa_{\mbox{\scriptsize sys}} \right)
\end{equation}


\subsubsection{Optical field} \label{sec:fov}

The optical field around IGR~J17091 is shown in Fig. \ref{fig:oqfc}, where we present small cut-outs from two good-seeing $r^{\prime}$-band images acquired on different nights. The target is in quiescence in the left-hand image (although a source is still visible at its location), and it is in outburst in the right-hand image (as the brightest source in the cut-out). As discussed in Sec.~\ref{sec:intro2}, IGR~J17091 has two (known) nearby sources, the closest at a separation of $\sim$0.4$^{\prime\prime}$ NE (unresolved and not marked in Fig.~\ref{fig:oqfc}; hereafter BS1, with ``BS'' standing for blend source), and the other at a separation of $\sim$1$^{\prime\prime}$ NNW (barely resolved and marked in Fig.~\ref{fig:oqfc} with a cyan circle; hereafter BS2). In our LCO images (with e.g. median seeing 1.9$^{\prime\prime}$ in the $r^{\prime}$-band), BS1 is always blended with IGR~J17091, while BS2 is blended in most of the images. This blending, along with any further hidden blend sources, will have introduced (seeing-dependent) systematic errors into our lightcurve of IGR~J17091, both for the multi-aperture photometry\footnote{Affected because the profile fits are performed without the subtraction of nearby sources.} and for the aperture photometry. Furthermore, the size of these systematic errors (in magnitudes) is a function of the brightness of IGR~J17091, with smaller errors when it is brighter, and larger errors when it is fainter. Any scientific analysis of the optical lightcurve therefore needs to account as best as possible for these errors.

To aid in this task, we searched for optical photometry of the two known blend sources. Photometric catalogues from ground-based surveys either do not cover the field of IGR~J17091 \citep[e.g. Pan-STARRS1 and DESI Legacy Imaging Survey;][]{2019AJ....157..168D} or do not have the resolution needed to separate the target and blend sources properly \citep[e.g. SDSS;][]{2000AJ....120.1579Y}. In contrast, from space, the \textit{Gaia} satellite has sufficient resolution to separate them \citep[$\sim$0.1$^{\prime\prime}$;][]{2016A&A...595A...3F}. Assuming that all three sources are also bright enough to be detected, then it follows that \textit{Gaia} DR3 should report (at least) three sources in the close vicinty of IGR~J17091. However, \textit{Gaia} DR3 reports only two sources within a radius of 3$^{\prime\prime}$ of the target coordinates (Sec.~\ref{sec:dataproc}). The closest \textit{Gaia} source to the search coordinates, at R.A. 17:09:07.610 and Dec. $-$36:24:25.51 (ICRS; reference epoch J2016), has a mean $G$-band magnitude of 20.08$\pm$0.03, and no proper motion measurement. The other \textit{Gaia} source, at R.A. 17:09:07.595 and Dec. $-$36:24:24.42 (ICRS; reference epoch J2016), has a mean $G$-band magnitude of 20.185$\pm$0.006, and a proper motion of $\sim$4.5~mas~yr$^{-1}$ (negligible for our purposes). This latter source is clearly BS2 when plotted in Fig.~\ref{fig:oqfc}. Given that \textit{Gaia} scanned the field containing IGR~J17091 on numerous occasions during its 2016 outburst when it was bright enough to be detected,\footnote{See https://gaia.esac.esa.int/gost/index.jsp} we may conclude that the \textit{Gaia} source closest to the search coordinates must be IGR~J17091 itself. Therefore, BS1 is undetected by \textit{Gaia} and must be fainter than the limiting magnitude $G\approx21$~mag. 

Unfortunately, we could not find any further space-based optical photometry of this field (e.g. HST, JWST, etc.). Hence, overall, we are limited in our prior knowledge about the blend sources in the optical. Specifically, we only have precise \textit{Gaia} coordinates for BS2, and we know that the brightnesses of the blend sources are $G>21$ and $G=20.185$~mag for BS1 and BS2, respectively.

\subsubsection{Deblending the optical lightcurve} \label{sec:dan}

To enable a valid analysis of the lightcurve of our target IGR~J17091, we must first isolate the flux at each epoch due solely to IGR~J17091 (i.e. deblend the lightcurve by removing the contaminating fluxes from the blend sources). This can be done either at the image processing stage, or as a post-processing step in the presence of sufficient information. We have opted for the latter, since XB-NEWS is not designed for the former. Furthermore, we find that while BS2 is visible in the $i^{\prime}$ and $r^{\prime}$-band images (e.g. Fig.~\ref{fig:oqfc}), it is not visible in any of our $V$-band images. Combining this with the fact that BS1 is even fainter than BS2 in the wider $G$-band, we believe that any flux contamination of the target lightcurve in the $V$-band will be negligible (certainly so during outburst). Hence, we do not attempt to deblend the $V$-band target lightcurve.

As the target and blend sources are unresolved in most of the LCO images, XB-NEWS typically detects them as a single blended source in an image, measures the centroid,\footnote{Technically, a windowed centroid as implemented in SExtractor.} and assigns the resulting coordinate estimates to the target lightcurve. We noticed that, for the $i^{\prime}$ and $r^{\prime}$-bands, the set of these coordinates in the target lightcurve cluster around (and sometimes between) the true locations of IGR~J17091 and BS2, with the clustering being around IGR~J17091 during outburst, and around BS2 during quiescence. This is easily explained as follows. When IGR~J17091 is in outburst, its flux dominates the total flux in the blend, and the centroid is shifted towards the true location of IGR~J17091. Conversely, when the target is in quiescence, the flux from BS2 dominates the total flux in the blend, and the centroid is shifted towards the true location of BS2. Since the measured coordinates in this case are estimates of the blend centroid, they also exhibit the same behaviour.

We used the target detections during outburst and across all wavebands to compute mean coordinates for IGR~J17091, which we report as the adopted/known target coordinates in Sec.~\ref{sec:dataproc}. As target coordinates are a required input for the XB-NEWS pipeline, the computation of mean coordinates and their subsequent use in generating the target lightcurve required two full runs of XB-NEWS on all of the reduced science images. Also, since the separation between the target and BS2 is $\sim$1$^{\prime\prime}$, noisy measurements of the blend centroid during quiescence are sometimes $>$1$^{\prime\prime}$ away from the target coordinates, and in these cases XB-NEWS adopts the target coordinates for the entry in the lightcurve (Sec.~\ref{sec:dataproc}). This also happens in cases of relatively poor atmospheric transparency (i.e. for non-detections).

All of the above means that the coordinates recorded in the target lightcurve, along with the image PSF FWHM, vary (substantially) between epochs. Since XB-NEWS performs photometry centred on these coordinates with seeing-matched profiles/apertures, the relative contributions of the target and blend sources to the measured total flux at each epoch also varies substantially. For the multi-aperture photometry, this situation is especially complicated because the profile fits are performed sequentially from small to large aperture radii coupled with an early stopping criterion to catch a poor fit or a lack of improvement in S/N. This severely impedes accurate modelling of how the blending affects the multi-aperture photometry in the target lightcurve. In contrast, it is simple to model the aperture photometry procedure in each image, and so we opt for this to enable subsequent deblending. 

We start by modelling the $i^{\prime}$ and $r^{\prime}$-band target lightcurves separately, and for epochs \textit{during quiescence} only.\footnote{Epochs with MJDs in the ranges 56700$-$57320, 57800$-$59600, and $>$59900.} We use the standard magnitude measurements $m_{\mbox{\scriptsize std},k}$ in the relevant waveband from the aperture photometry with the aperture radius of 0.75$\times$ the image PSF FWHM, and convert them (back) to fluxes $f^{\prime}_{\mbox{\scriptsize std},k}$ using a magnitude zero-point of 25. The index $k$ is for the $k$th image. These fluxes are related to $f_{\mbox{\scriptsize std},k}$ (equation~(3)) by a constant factor:
\begin{equation}
f^{\prime}_{\mbox{\scriptsize std},k} = \phi f_{\mbox{\scriptsize std},k}
\end{equation}
where:
\begin{equation}
\phi = 10^{\, 0.4 \left( 25 - z_{\mbox{\tiny std}} \right)}
\end{equation}
Using the fluxes $f^{\prime}_{\mbox{\scriptsize std},k}$ as data instead of $f_{\mbox{\scriptsize std},k}$ avoids the need to compute $z_{\mbox{\scriptsize std}}$ in any of the following.

Our model for the data $f^{\prime}_{\mbox{\scriptsize std},k}$ first assumes that IGR~J17091, BS1, and BS2 all have (unknown) constant fluxes $f_{\mbox{\scriptsize std,tar}}$, $f_{\mbox{\scriptsize std,BS1}}$, and $f_{\mbox{\scriptsize std,BS2}}$, respectively, at the top of the atmosphere, which are attenuated by the factor $\kappa_{\mbox{\scriptsize sys},k}$ in the $k$th image. Then we assume that the $k$th image $I_{k}(x,y)$ has a Gaussian PSF with FWHM equal to the value estimated by XB-NEWS for the image (Sec.~\ref{sec:dataproc}). From this we obtain a normalised model PSF $P_{k}(u,v)$ for $I_{k}$. Writing the known \textit{Gaia} coordinates for IGR~J17091 and BS2 in $I_{k}$ as $(x_{\mbox{\scriptsize tar},k}, y_{\mbox{\scriptsize tar},k})$ and $(x_{\mbox{\scriptsize BS2},k}, y_{\mbox{\scriptsize BS2},k})$, respectively, we can then model the normalised flux profiles for IGR~J17091 and BS2 in $I_{k}$ as \mbox{$P_{k}(x - x_{\mbox{\scriptsize tar},k}, \, y - y_{\mbox{\scriptsize tar},k})$} and \mbox{$P_{k}(x - x_{\mbox{\scriptsize BS2},k}, \, y - y_{\mbox{\scriptsize BS2},k})$}, respectively. This is not possible for BS1 in a consistent way as it lacks \textit{Gaia} coordinates. However, with BS1 being fainter than both BS2 and \textit{Gaia}'s limiting magnitude, failing to remove the contaminating flux of BS1 from the target lightcurve will only leave small systematic errors in the magnitudes near quiescence, while hardly affecting the magnitudes during outburst. Hence, we do not consider BS1 any further in the deblending analysis.

Consider now the aperture used for the aperture photometry of the target in $I_{k}$, which is centred at the coordinates recorded in the target lightcurve. Using numerical techniques, we integrate the normalised flux profiles \mbox{$P_{k}(x - x_{\mbox{\scriptsize tar},k}, \, y - y_{\mbox{\scriptsize tar},k})$} and \mbox{$P_{k}(x - x_{\mbox{\scriptsize BS2},k}, \, y - y_{\mbox{\scriptsize BS2},k})$} over this aperture to obtain the fractions $\omega_{\mbox{\scriptsize tar},k}$ and $\omega_{\mbox{\scriptsize BS2},k}$, respectively. Putting everything that we have so far together, we may write the following expression for the model total instrumental flux $f_{\mbox{\scriptsize ins,tot},k}$ from IGR~J17091 and BS2 in $I_{k}$ that falls within the photometric aperture:
\begin{equation}
f_{\mbox{\scriptsize ins,tot},k} = \kappa_{\mbox{\scriptsize sys},k} \left( \, \omega_{\mbox{\scriptsize tar},k} \, f_{\mbox{\scriptsize std,tar}} + \omega_{\mbox{\scriptsize BS2},k} \, f_{\mbox{\scriptsize std,BS2}} \, \right)
\end{equation}
The corresponding zero-point offset $\Delta z_{k}$ from the fitted photometric calibration model (Sec.~\ref{sec:dataproc}) is dependent on the typical value of $\omega_{\mbox{\scriptsize apr},k}$ for sources in $I_{k}$ (equation~(5)). By assuming that the aperture is correctly centred at the peak of the flux profile for the vast majority of the sources in $I_{k}$, we may compute $\omega_{\mbox{\scriptsize apr},k}$ as the integral of $P_{k}(u,v)$ from the origin out to the aperture radius, yielding \mbox{$\omega_{\mbox{\scriptsize apr},k} = 0.7896$} for all images. Then, substituting equation~(8) into equation~(4), multiplying both sides by $\phi$, and rearranging, we obtain the model total flux from IGR~J17091 and BS2 in $I_{k}$:
\begin{equation}
f^{\prime}_{\mbox{\scriptsize std,tot},k} = \left( \frac{ \omega_{\mbox{\scriptsize tar},k} }{ \omega_{\mbox{\scriptsize apr},k} } \right) \phi f_{\mbox{\scriptsize std,tar}} + \left( \frac{ \omega_{\mbox{\scriptsize BS2},k} }{ \omega_{\mbox{\scriptsize apr},k} } \right) \phi f_{\mbox{\scriptsize std,BS2}}
\end{equation}
which is the model flux for our flux data $f^{\prime}_{\mbox{\scriptsize std},k}$.

Assuming independent Gaussian errors on $f^{\prime}_{\mbox{\scriptsize std,tot},k}$ as determined by XB-NEWS, we fit the model in equation~(9) to the flux data $f^{\prime}_{\mbox{\scriptsize std},k}$. We do this by optimising the values of the free parameters \mbox{$\gamma = \phi f_{\mbox{\scriptsize std,tar}}$} and \mbox{$\delta = \phi f_{\mbox{\scriptsize std,BS2}}$} to minimise the chi-squared, which yields best-fit parameter values $\hat{\gamma}$ and $\hat{\delta}$.\footnote{The best-fit parameter values $\hat{\gamma}$ and $\hat{\delta}$ may be computed analytically given that equation~(9) is a linear model in $\gamma$ and $\delta$.} Interestingly, while the best-fit solution is potentially degenerate, it is not in this case specifically because of the substantial variations in the fractions $\omega_{\mbox{\scriptsize tar},k}$ and $\omega_{\mbox{\scriptsize BS2},k}$ between epochs. From equations~(3)~and~(7), we then compute the (mean) standard magnitude of IGR~J17091 in quiescence \mbox{$m_{\mbox{\scriptsize std,tar}} = 25 - 2.5 \log_{\mbox{\scriptsize 10}} \left( \gamma \right)$} using $\gamma = \hat{\gamma}$ as $\sim$20.992$\pm$0.020 and $\sim$21.817$\pm$0.003 in the $i^{\prime}$ and $r^{\prime}$-bands, respectively. Similarly, we compute the (mean) standard magnitude of BS2 \mbox{$m_{\mbox{\scriptsize std,BS2}} = 25 - 2.5 \log_{\mbox{\scriptsize 10}} \left( \delta \right)$} using $\delta = \hat{\delta}$ as $\sim$19.645$\pm$0.005 and $\sim$21.062$\pm$0.001 in the $i^{\prime}$ and $r^{\prime}$-bands, respectively.

With our flux estimate $\hat{\delta}$ for BS2, the deblending of the target lightcurve can now be done as follows. For \textit{any epoch} in the target lightcurve, we may write the following expression, derived in a similar manner to equation~(8):
\begin{equation}
f_{\mbox{\scriptsize ins,tot},k} = \kappa_{\mbox{\scriptsize sys},k} \left( \, \omega_{\mbox{\scriptsize tar},k} \, f_{\mbox{\scriptsize std,tar},k} + \omega_{\mbox{\scriptsize BS2},k} \, f_{\mbox{\scriptsize std,BS2}} \, \right)
\end{equation}
The difference between this expression for $f_{\mbox{\scriptsize ins,tot},k}$ and the one in equation~(8) is that the target flux $f_{\mbox{\scriptsize std,tar},k}$ is now allowed to be a different (unknown) value at each epoch. From equations~(1)~and~(2), we may also write:
\begin{equation}
10^{\, 0.4 \Delta z_{k}} f_{\mbox{\scriptsize ins,tot},k} = 10^{\, 0.4 \left( 25 - m_{\mbox{\tiny std,tot},k} \right)}
\end{equation}
where $m_{\mbox{\scriptsize std,tot},k}$ is the model standard magnitude corresponding to $f_{\mbox{\scriptsize ins,tot},k}$. Taking equations~(3), (5), (7), (10)~and~(11), along with the definition of $\delta$, and then doing some algebra, yields:
\begin{equation}
m_{\mbox{\scriptsize std,tar},k} = 25 - 2.5 \log_{\mbox{\scriptsize 10}} \left( \, \left( \frac{ \omega_{\mbox{\scriptsize apr},k} }{ \omega_{\mbox{\scriptsize tar},k} } \right) 10^{\, 0.4 \left( 25 - m_{\mbox{\tiny std,tot},k} \right)} - \left( \frac{ \omega_{\mbox{\scriptsize BS2},k} }{ \omega_{\mbox{\scriptsize tar},k} } \right) \delta \, \right)
\end{equation}
where $m_{\mbox{\scriptsize std,tar},k}$ is the standard (deblended) magnitude of IGR~J17091 in the $k$th image. The deblended target lightcurve then consists of all $m_{\mbox{\scriptsize std,tar},k}$ estimates as obtained from equation~(12) by using the computed values for $\omega_{\mbox{\scriptsize tar},k}$, $\omega_{\mbox{\scriptsize BS2},k}$, and $\omega_{\mbox{\scriptsize apr},k}$, setting \mbox{$\delta = \hat{\delta}$}, and using the standard magnitude measurements $m_{\mbox{\scriptsize std},k}$ from the (original) target lightcurve as estimates of $m_{\mbox{\scriptsize std,tot},k}$. The magnitude uncertainties in the deblended target lightcurve are obtained from the uncertainties provided by XB-NEWS for $m_{\mbox{\scriptsize std},k}$ by updating them using standard (first order) error propagation as applied to equation~(12).

After processing all of the reduced science images for IGR~J17091 with XB-NEWS, removing the 6 unusable $r^{\prime}$-band images, applying the de-blending procedure described above, and performing a cut to exclude all points fainter than magnitude 22, we obtain a lightcurve with 409, 384, and 55 epochs in the $i^{\prime}$, $r^{\prime}$, and $V$-bands, respectively, that spans the time period 2011 April 26 (MJD 55677) to 2024 Oct 27 (MJD 60610).

\subsection{Infrared monitoring} \label{sec:ir}

\subsubsection{Mid-infrared observations}

Mid-IR observations of the field of IGR~J17091 were made with the Very Large Telescope (VLT) on 2016 March 20, under the program 096.D-0467 (PI: D. Russell). The VLT Imager and Spectrometer for the mid-Infrared \citep[VISIR;][]{LagageVISIR} instrument on the VLT was used in small-field imaging mode (45 mas pixel$^{-1}$). Observations were made in two filters, J8.9 (central wavelength 8.70 $\mu$m; at 09:21--09:40 UT; MJD 57467.40) and PAH2$\_$2 (11.68 $\mu$m; at 09:41--10:07 UT; MJD 57467.41). Conditions were good, with a moderate precipitable water vapor column (PWV $\sim$6 mm) and photometric conditions according to standard star observations taken directly after the target (thus the target observations were not affected by PWV). For each observation, the integration time on source consisted of a number of nodding cycles, with chopping and nodding between source and sky. The total observing time is typically almost twice the integration time. After the final chop/nod combinations were applied, the effective VISIR field of view is $\sim$19.2 arcsec. Although the raw frames span a larger area, the overlap between the positive and negative beams from the chop/nod pattern reduces the final science-ready FOV to this value.

Observations of mid-IR standard stars were made on the same night as the IGR~J17091, in the same filters. HD099167 was observed in PAH2$\_$2 at 05:18 UT (MJD 57467.22) before the target observations, and HD178345 was observed in J8.9 at 10:12 UT and PAH2$\_$2 at 10:14 UT (MJD 57467.43) immediately after IGR~J17091. All observations (target and standard stars) were reduced using the VISIR pipeline in the \emph{gasgano} environment, and raw images from the chop/nod cycle were combined. Similarly to other sources \citep[e.g.]{Baglio2018,saikia1716,constanza}, photometry was performed on the combined images using an aperture size of 0.75 arcsec, which is large enough that small seeing variations did not affect the fraction of flux in the aperture. The counts/flux ratio values were calculated from the standard stars. From the two standards taken before and after the target, the counts/flux ratio differed by 2.1 per cent, confirming that the conditions were photometric. The airmass of the target was 1.03, and the standards were both taken at an airmass of 1.12. The overall long-term and night-to-night stability of the photometric calibration of VISIR is known to be good \citep{Dobrzycka12}, and any variations due to changing airmass and visibility conditions were minimal on the night.

No clear source was visible in the combined VISIR images of IGR~J17091. The brightest possible point source appeared in the J8.9 filter image, with a flux density of 7.7 $\pm$ 1.3 mJy and a signal-to-noise ratio (S/N) of 6.6. However, it is uncertain whether this is the counterpart to IGR~J17091. This is in a crowded region of the Galactic plane, and several slightly less significant potential sources appear in the J8.9 and PAH2$\_$2 images. None of the potential sources appear at the same location in both images, and most have S/N $<$ 4. We therefore conclude that IGR~J17091 was not unambiguously detected, and by sampling the background, we derive $3\sigma$ flux density upper limits of 2.4 mJy in J8.9 and 4.9 mJy in PAH2$\_$2.

\subsubsection{Near-infrared monitoring}

We also compiled infrared photometric data for IGR~J17091 from multiple observing campaigns that span quiescence and outburst states, to build quasi-simultaneous optical/infrared spectra. In quiescence, \cite{Torres2011} reported a Ks-band magnitude of 16.98 $\pm$ 0.04 (Vega) on 2008 June 23 (MJD 54640), with the target deblended from BS1 using PSF fitting. During the 2016 outburst, \cite{grond} obtained optical and infrared magnitudes under 1.3 arcsec seeing, where the target was blended with both BS1 and BS2. The observed values in the AB magnitudes were $g^{\prime}$ = 22.5 $\pm$ 0.1 mag, $r^{\prime}$ = 20.2 $\pm$ 0.1 mag, $i^{\prime}$ = 19.2 $\pm$ 0.1 mag, $z^{\prime}$ = 18.5 $\pm$ 0.1 mag, $J$ = 17.4 $\pm$ 0.1 mag, $H$ = 17.1 $\pm$ 0.1 mag, and $K_s$ = 17.0 $\pm$ 0.1 mag. More recently, \cite{john} used an image subtraction technique to isolate the source flux in WISE W1 and W2 bands, removing contamination from BS1 and BS2. These observations revealed NEOWISE detections on three days - once from the 2016 outburst and twice in the 2022 outburst. We use these values in Section \ref{fig:sed} to build the optical/infrared SEDs.

\subsection{X-ray monitoring} \label{sec:obsxray}
 
\subsubsection{\emph{Swift}/XRT}
IGR~J17091 was routinely monitored by the X-Ray Telescope \citep[XRT;][]{burrows} onboard the Neil Gehrels \emph{Swift} observatory during its outbursts. In this study, we used all available \emph{Swift}/XRT observations of the source during the 2011, 2018 and 2022 outbursts. The light-curves were retrieved using the online \emph{Swift}/XRT data products generator\footnote{\url{https://www.swift.ac.uk/user objects/}} maintained by the \emph{Swift} data center at the University of Leicester \citep[see][]{evans1,evans2}. We extracted the $2-10$ keV light curve, as well as the $1.5-10$ keV/$0.6-1.5$ keV X-ray hardness ratio, after binning the data by observation, and removed all the data that has S/N < 3 for quality control.

\begin{figure*}
    \centering
    \includegraphics[width = 2.0\columnwidth]{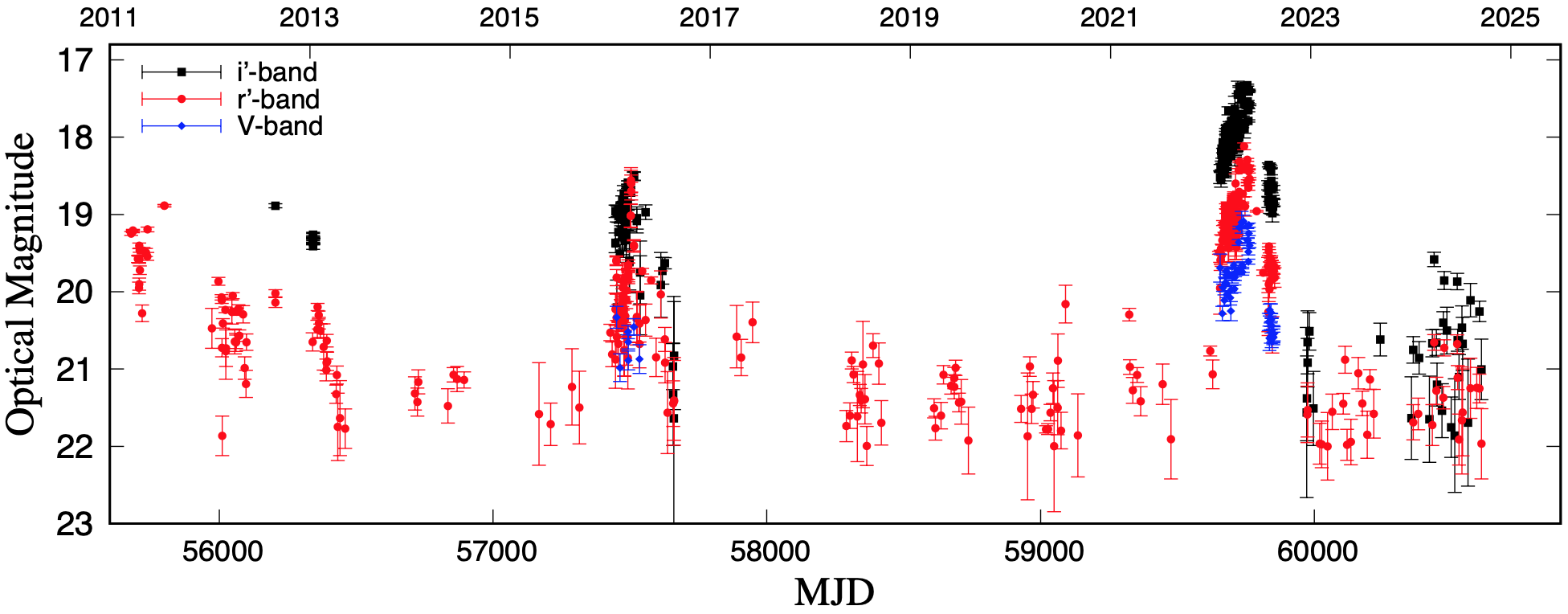}
    \caption{Long-term optical lightcurve of IGR~J17091 covering the 2011, 2016 and 2022 outbursts in $i^{\prime}$ (black squares), $r^{\prime}$ (red circles) and V-band (blue diamonds) with the LCO telescopes. For zoomed-in versions of the light-curve during outbursts, please see Fig. \ref{fig:mlc}.}
    \label{fig:opt_complete}
\end{figure*}

\begin{figure}
    \centering
    \includegraphics[width = \columnwidth]{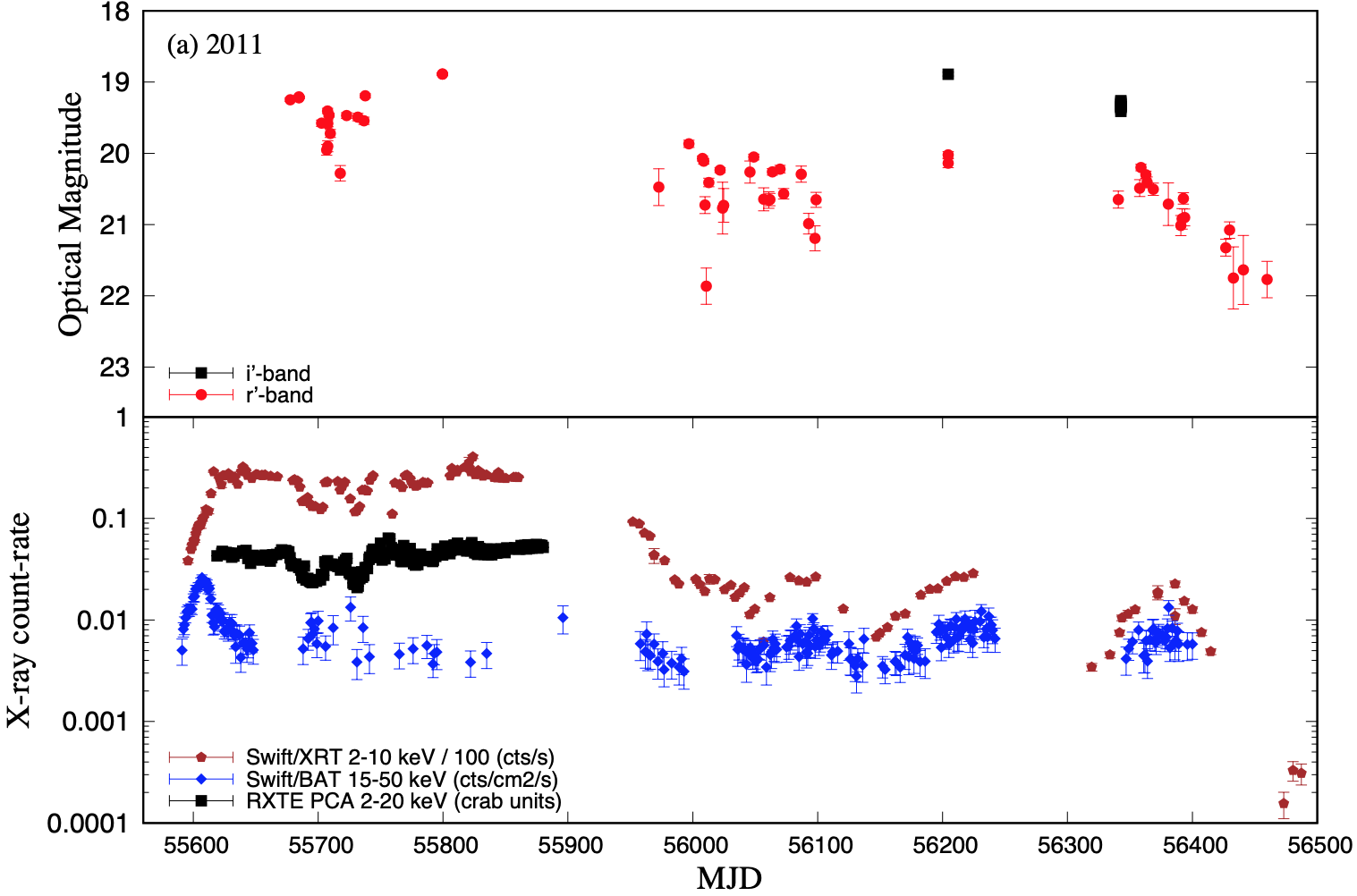}
    \includegraphics[width = 1.0001\columnwidth]{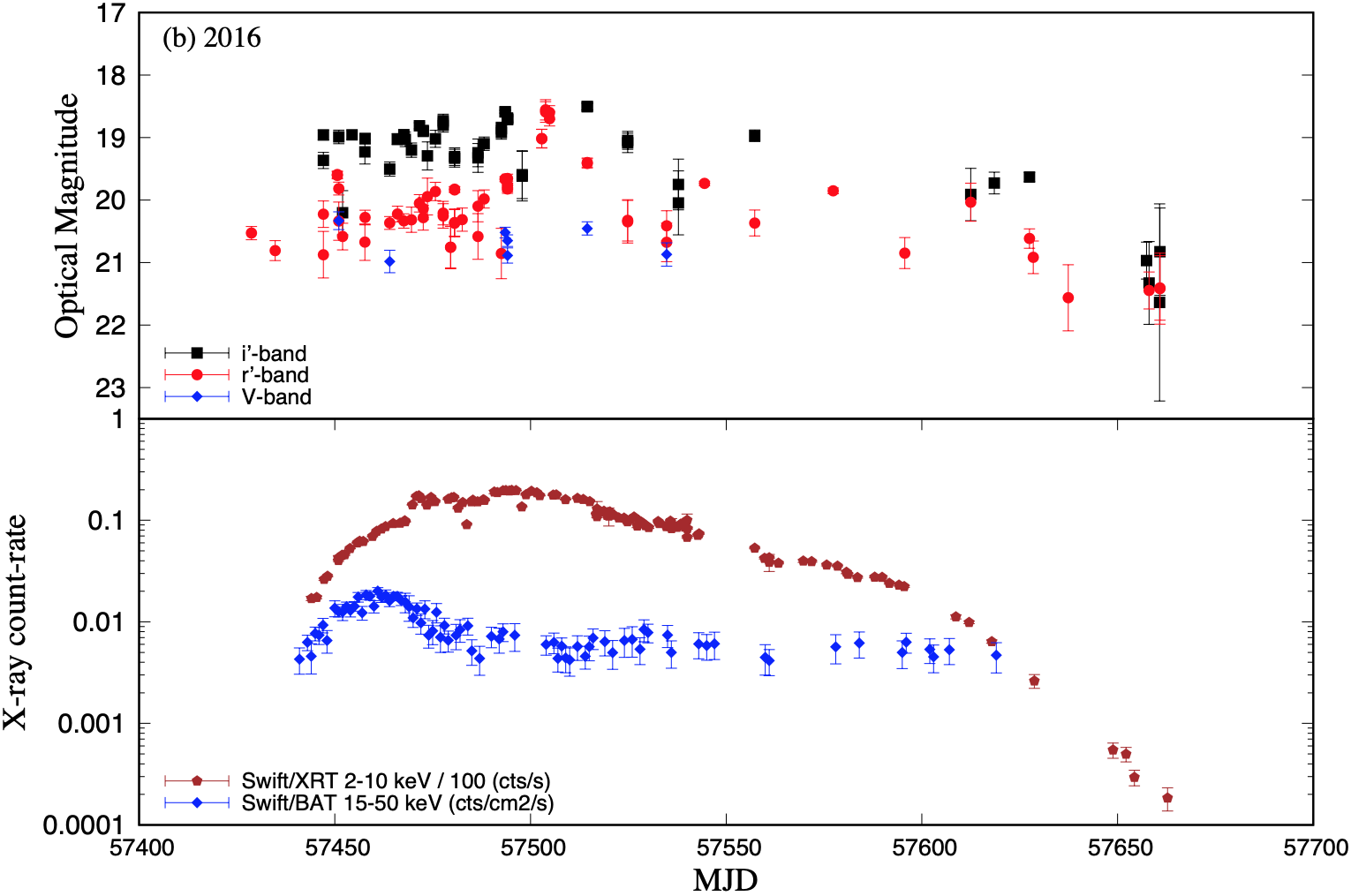}
    \includegraphics[width = \columnwidth]{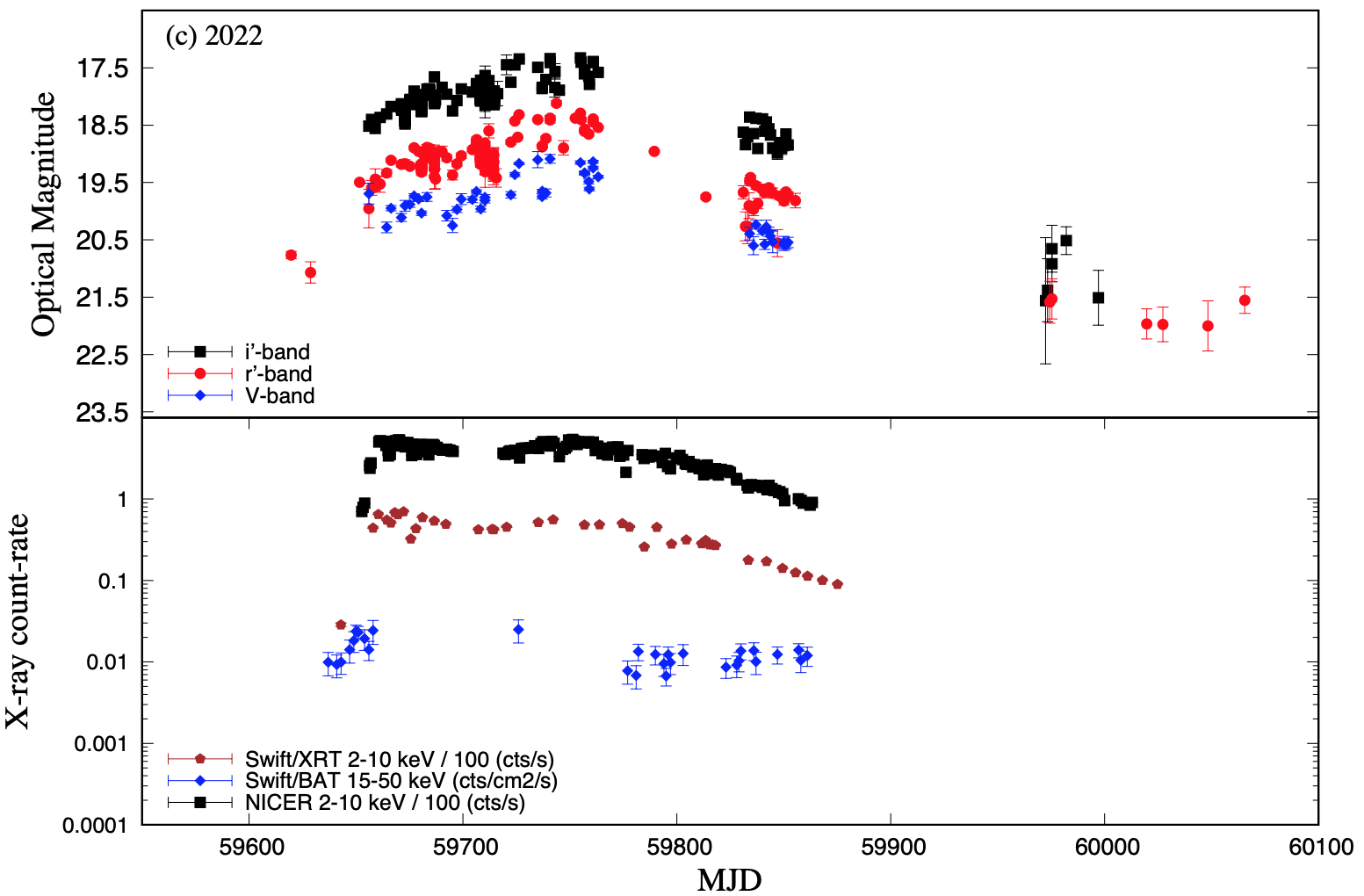}
    \caption{Zoom in of the optical and X-ray lightcurves from the last three outbursts of IGR~J17091 in 2011 (top panel), 2016 (middle panel) and 2022 (bottom panel), respectively. Note that the X-ray count rates for different instruments/telescopes are shown in different units (specified in the legends) with Swift/XRT and NICER count rates given in cts/s, Swift/BAT in cts/cm2/s and RXTE in crab units.}
    \label{fig:mlc}
\end{figure}

\subsubsection{\emph{Swift}/BAT}
IGR~J17091 was observed by the all-sky monitor on board the \emph{Swift} observatory, known as the Burst Alert Telescope \citep[BAT;][]{krimm}, which observes the hard X-ray sources in the sky including BHXBs. We gathered the \emph{Swift}/BAT 15-50 keV X-ray light curve of IGR~J17091 with a one-day time bin, from the \emph{Swift}/BAT archive\footnote{\url{https://swift.gsfc.nasa.gov/results/transients}} \citep[][]{krimm}. For all the available \emph{Swift} X-ray observations, we only consider the data with S/N > 3 as proper detections, and remove the rest from further analysis.

\subsubsection{RXTE}
IGR~J17091 was also monitored almost every day with the Proportional Counter Array (PCA) on-board RXTE \citep{zhang} during its 2011 outburst. We used the RXTE $2-20$ keV X-ray lightcurve from \cite{diego2011}, produced using standard reduction techniques \citep[e.g.][]{bell}. Due to contamination from the bright and variable source GX 349+2 (Sco X-2), which was located within the 1\degr \, PCA field of view, the first 10 observations of IGR~J17091 taken with RXTE are affected \citep[see also][]{diego2011,rodri}. As a result, we exclude all RXTE observations of the source up to MJD 55615 from our analysis.

\subsubsection{NICER}
The Neutron Star Interior Composition Explorer \citep[\textit{NICER};][]{gen} observed IGR~J17091 during the 2022 outburst 175 times from 2022 March 14 (ObsID $5202630101$) to 2022 Oct 12 (ObsID $5618011453$). We study in detail the seven observations that were quasi-simultaneous (i.e., same MJD) with LCO (ObsIDs $5202630112$, $4618020801$, $5202630114$, $5202630116$, $5202630119$, $5202630123$ and $5202630125$). We analyze the data using the software HEASOFT version 6.33, NICERDAS 11.0 and CALDB 20221001, applying standard filtering and cleaning criteria. We include data when the dark Earth limb angle was $>15\degr$, the pointing offset was $<54\arcsec$, the bright Earth limb angle was $>30\degr$, and the International Space Station was outside the South Atlantic anomaly. Moreover, we remove data from detectors 14 and 34 since they show episodes of electronic noise. We correct the event times to the Solar system barycenter time using the \textsc{barycorr} tool with the coordinates R.A.$=$17:09:07.605 and Dec.$=-$36:24:25.35, and create 1s, 2s, and 120s binned light curves for each ObsID in the 0.5--10.0 keV energy band.

We also obtain the PDS of the seven observations with \textit{NICER}. We construct the Leahy-normalized \citep[][]{Leahy83} PDS using data segments of 262.14 seconds and a time resolution of 250$\mu$s. The minimum frequency is $\sim$0.004 Hz, and the Nyquist frequency is 2000 Hz. We then average the PDS per ObsID and subtract the Poisson noise based on the average power in the $500-2000$ Hz frequency range. Finally, we normalize the PDS to fractional rms \citep[][]{bh90}.

\section{Results} \label{sec:results}

\subsection{Multi-wavelength light curves} \label{sec:lc}

We present optical monitoring data of IGR~J17091 from our long term LCO campaign from 2011 April 26 (MJD 55677) to 2024 Oct 27 (MJD 60610). As shown in Fig. \ref{fig:opt_complete}, these data encompass the 2011, 2016, and 2022 outbursts in the $V$, $i^{\prime}$, and $r^{\prime}$-bands, as well as the intervening quiescent periods primarily in the $r^{\prime}$-band. 
During the outbursts, the three optical bands exhibit correlated behavior. To characterize the time evolution of optical and X-ray flux, we present detailed light curves for each of the three outbursts in Fig. \ref{fig:mlc}.

\begin{figure*}
    \centering
    \includegraphics[width = 0.976\columnwidth]{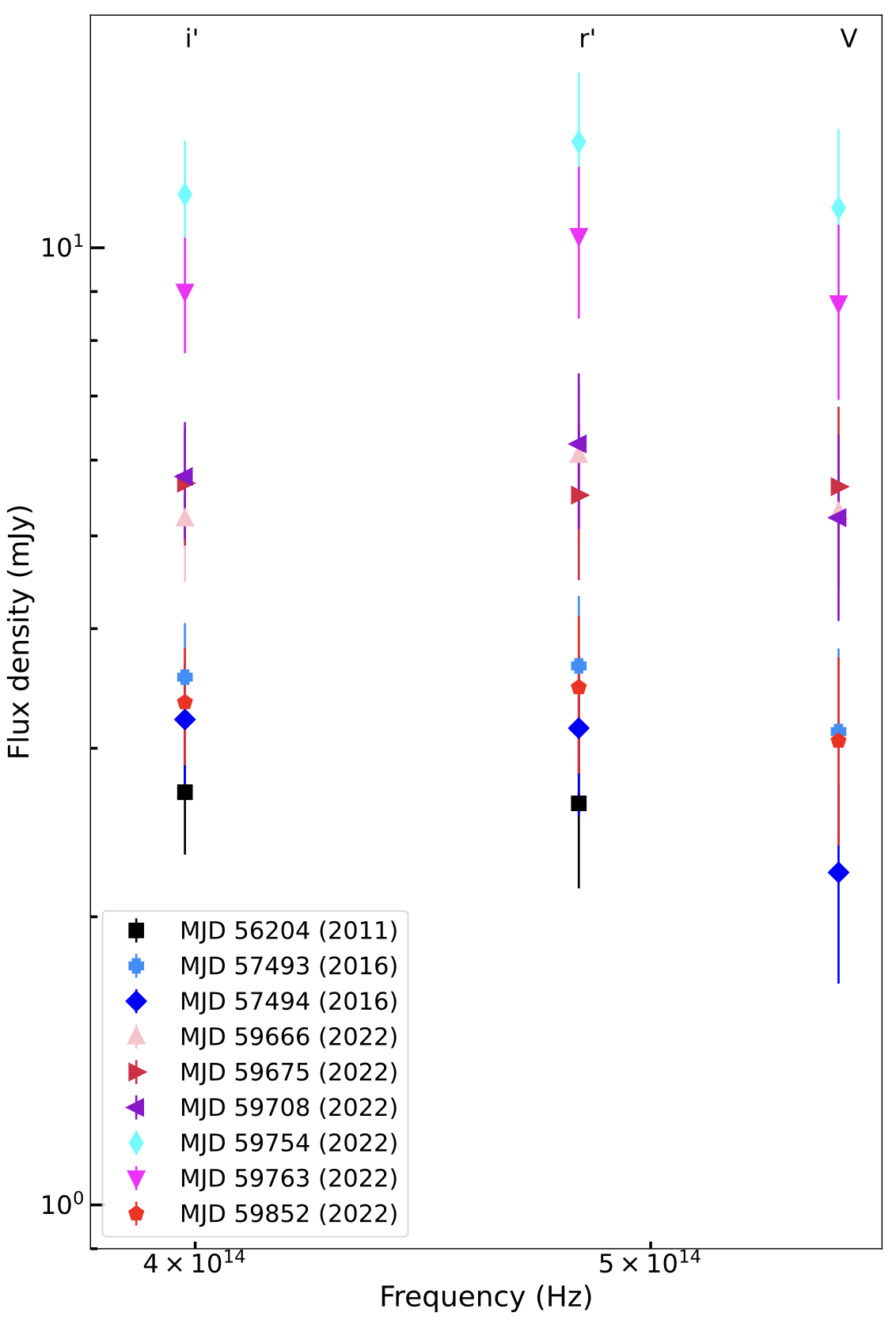}
    \includegraphics[width = 1.02\columnwidth]{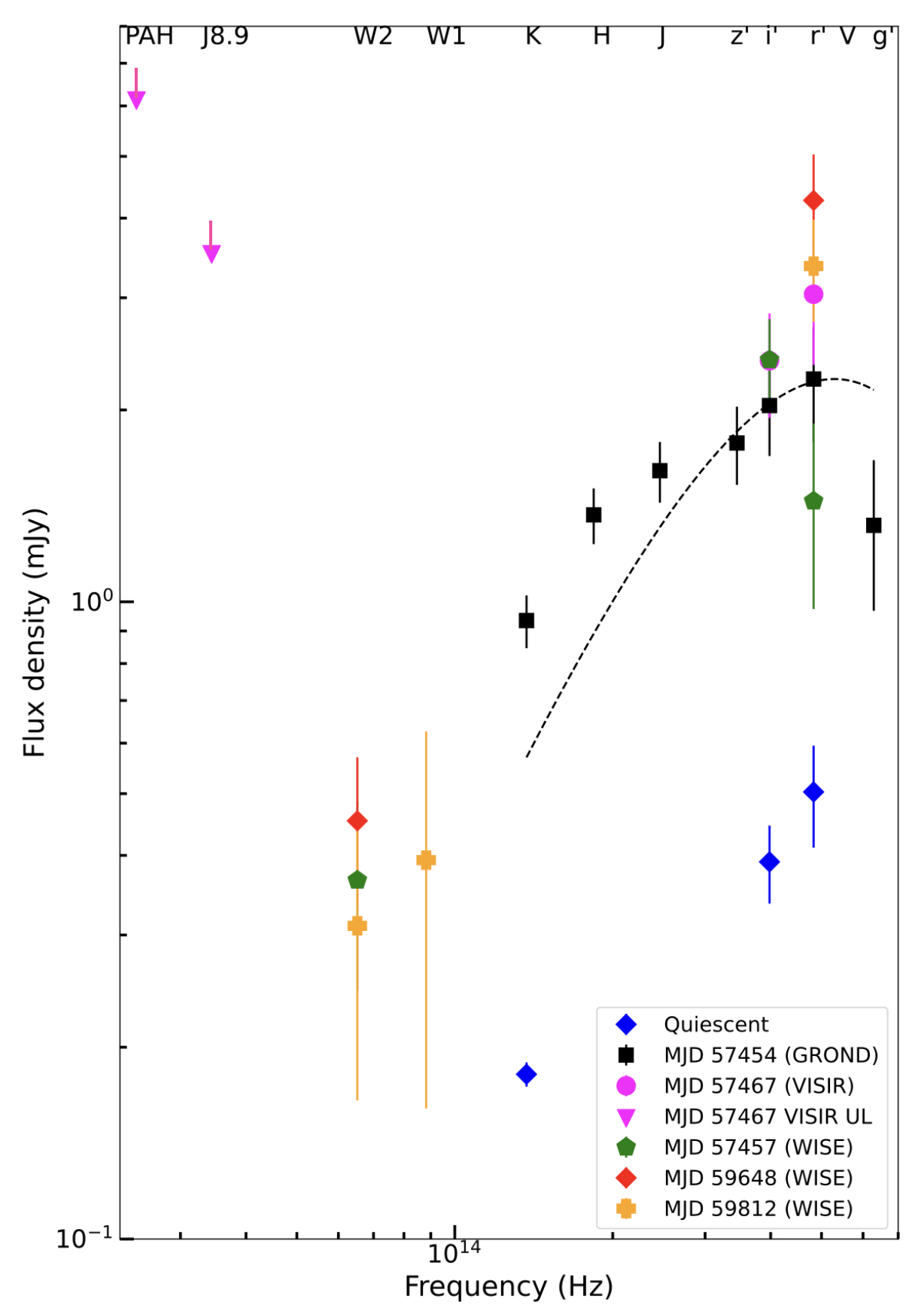}
    \caption{(Left panel) De-reddened optical spectral energy distributions of IGR~J17091 during the three outbursts, assuming $N_{\mbox{\scriptsize H}}=(1.537\pm0.002)\times 10^{22}\, \rm cm^{-2}$ or $A_{V} = 5.36\pm0.22$ mag \citep{wang2}. (Right panel) De-reddened optical/IR SED of IGR~J17091 including avavilable data in the NIR and upper limits in the mid-IR ranges, using the same $N_{\mbox{\scriptsize H}}$ value. The quiescent SED combines the $K_s$-band measurement from Magellan/PANIC \citep{Torres2011}, deblended from the nearby sources BS1 and BS2, with the mean quiescent $i'$ and $r'$-band magnitudes from our LCO observations.} We also overlay a blackbody fit to the GROND data (which has the most quasi-simultaneous data points) to achieve the best alignment with the optical data (black dashed line, T = 9000 K).
    \label{fig:sed}
\end{figure*}

The 2011 outburst began around 2011 Feb 8 (MJD 55600), as seen in the initial rise of the Swift/XRT (2–10 keV) and RXTE PCA (2–20 keV) X-ray flux, reaching peak levels shortly after (see Fig. \ref{fig:mlc}a). The optical observations started slightly after the X-ray peak was reached, with the peak optical magnitude recorded as $r^{\prime}$= 19.22$\pm$0.01 on 2011 May 3 (MJD 55684). The system maintained a relatively high X-ray flux until around 2011 Dec 5 (MJD 55900), after which a gradual decay phase set in, with the soft X-ray (Swift/XRT, 2--10 keV) flux declining significantly. The hard X-ray emission (Swift/BAT, 15--50 keV) remained variable but followed a similar downward trend. Optical magnitudes also faded over this period. Re-brightening events occured around 2012 Aug 11 (MJD 56150) and then again in 2013 Feb 27 (MJD 56350), evident in both soft and hard X-ray bands \citep[see \ref{fig:mlc}a, as also noted by][]{pereyra}, before the system finally faded to quiescence by 2013 July 27 (MJD 56500). Direct comparison of the X-ray reflares with optical data is challenging due to the sporadic nature of LCO observations during the re-brightening phases.

The 2016 outburst of IGR J17091, shown in Fig. \ref{fig:mlc}b, began around 2016 Feb 22 (MJD 57440), with a steady rise in the soft X-rays (Swift/XRT, 2–10 keV). The hard X-ray emission (Swift/BAT, 15–50 keV) followed a similar trend but exhibited greater variability during the rising phase. In the optical, the $V$, $i^{\prime}$, and $r^{\prime}$-band magnitudes also brightened as the outburst progressed. The peak optical magnitude recorded was $r^{\prime}$= 18.59$\pm$0.16 on 2016 April 25 (MJD 57503), which is almost 0.6 magnitude brighter than the 2011 peak magnitude. After reaching the peak, the system underwent a gradual decay in both X-ray and optical bands, with the system returning to quiescence by 2016 Oct 4 (MJD 57665). No re-brightening events were observed during or after their decay to quiescence.

Finally, the 2022 outburst of IGR~J17091 (see Fig. \ref{fig:mlc}c) began around 2022 March 22 (MJD 59660), with a steady rise in both the X-rays and optical. The optical peak was reached around 2022 June 24 (MJD 59754), with peak magnitude $r^{\prime}$= 18.29$\pm$0.02, which is brighter than both previous outbursts. The system maintained a high flux state until approximately 2022 Aug 9 (MJD 59800), after which the soft X-ray flux began a gradual decline, accompanied by a decrease in optical brightness. The hard X-ray flux remained low and variable throughout the decay phase. By 2022 Nov 17 (MJD 59900), the system had significantly faded in both X-ray and optical bands. 

The overall optical evolution of all three outbursts follows a similar trend, showing correlated optical and soft X-ray behavior throughout the outburst cycles.

\subsection{Spectral energy distribution} \label{sec:sed}

At optical frequencies, various components contribute to the emission of a BHXB. To identify the dominant mechanism behind the observed optical emission, we construct the de-reddened spectral energy distributions (SEDs) for IGR~J17091 (see Fig. \ref{fig:sed}). These SEDs are derived from quasi-simultaneous observations (within a 24-hour period) collected across different outbursts.

\subsubsection{Optical SED}

The optical fluxes are first de-reddened using the most commonly cited hydrogen column density for the source, $N_{\mbox{\scriptsize H}}=(1.1\pm0.3)\times 10^{22}\, \rm cm^{-2}$ \citep{capi,rodri,krimm}. Using the \citet{Foight16} relation, we estimate the extinction in the $V$-band ($A_{V}$) from $N_{\mbox{\scriptsize H}}$, resulting in $A_{V}\sim 3.83\pm1.05$ mag, and apply the \citet{Cardelli1989} extinction laws to calculate the extinctions for other optical bands. To compare the optical spectral indices across different outbursts and understand the emission origin, we fit the SED with the function $S_{\nu} \propto \nu^{\alpha}$, where $S_{\nu}$ is the de-reddened flux density, $\nu$ is the frequency and $\alpha$ is the spectral index. 

We find that the optical spectra obtained using $A_{V} = 3.83\pm1.05$ mag are quite red, with slopes ranging from $-1.0$ to $-2.0$, during all three outbursts. Such steep spectral indices are not physically expected for BHXBs in outburst (see Sections \ref{sec:emission} and \ref{sec:extinction} for a detailed discussion), although the large uncertainty in $A_{V}$, and consequently in the de-reddened fluxes, leads to correspondingly large uncertainties on the derived spectral indices. Therefore, to obtain robust uncertainties on $\alpha$, we performed the SED fittings using a Monte Carlo approach. We generated $10^{5}$ realisations of the de-reddened fluxes in each of the three filters by randomly drawing from normal distributions of the measured magnitudes and $A_{V}$ values. Each of these SEDs was then fitted with $S_{\nu} \propto \nu^{\alpha}$ using least-squares, and the median optical spectral indices, as well as the uncertainties, were derived from the resulting distribution of best-fit values (see Table \ref{table:sed}).


\begin{table}
\centering
\caption{List of optical spectral indices ($\alpha_{V-{i}^{\prime}}$, unless stated otherwise) for a few epochs analysed in this work. The indices derived using the commonly adopted extinction value $N_{\mbox{\scriptsize H}}=(1.1\pm0.3)\times10^{22}{\rm cm^{-2}}$ are listed in the second and third columns, while those obtained using the most recent extinction measurement $N_{\mbox{\scriptsize H}}=(1.537\pm0.002)\times10^{22}{\rm cm^{-2}}$ are given in the fourth and fifth columns. Uncertainties were estimated via $10^{5}$ Monte Carlo realisations of the fluxes. The quoted spectral indices correspond to the medians of the resulting distributions, with the $-1\sigma$ and $+1\sigma$ uncertainties defined using the 16th and 84th percentiles, respectively.}
\label{table:sed}
\begin{tabular}{l c c c c}
\hline
\hline
MJD 
& $\alpha_{1.1e22}$ 
& $( -\sigma, +\sigma )$
& $\alpha_{1.5e22}$
& $( -\sigma, +\sigma )$ \\
\hline

\textbf{2011} \\
56204 & $-1.64^{a}$ & -- & $-0.14^{a}$ & -- \\

\hline
\textbf{2016} \\
57493 & $-1.71$ & $(-0.98,+0.97)$ & $-0.34$ & $(-0.28,+0.29)$ \\
57494 & $-2.00$ & $(-0.95,+0.93)$ & $-0.72$ & $(-0.33,+0.34)$ \\

\hline
\textbf{2022} \\
59666 & $-1.22$ & $(-0.95,+0.93)$ & $+0.12$ & $(-0.24,+0.24)$ \\
59675 & $-1.47$ & $(-1.03,+1.02)$ & $-0.00$ & $(-0.26,+0.27)$ \\
59708 & $-1.56$ & $(-0.96,+0.94)$ & $-0.22$ & $(-0.24,+0.24)$ \\
59754 & $-1.37$ & $(-0.94,+0.92)$ & $-0.04$ & $(-0.21,+0.21)$ \\
59763 & $-1.34$ & $(-0.94,+0.92)$ & $-0.01$ & $(-0.20,+0.20)$ \\
59852 & $-0.99$ & $(-0.96,+0.95)$ & $+0.33$ & $(-0.31,+0.32)$ \\
\hline
\end{tabular}

\par
$^{a}$ Estimating $\alpha_{{r}^{\prime}-{i}^{\prime}}$ due to the absence of $V$-band data.
\end{table}



To investigate further, we reconstructed the optical spectra using the most recent value of $N_{\mbox{\scriptsize H}}$ obtained from X-ray studies, $N_{\mbox{\scriptsize H}}=(1.537\pm0.002)\times 10^{22}\, \rm cm^{-2}$, which translates into an extinction of $A_{ V}\sim 5.36\pm0.22$ mag \citep{wang2}. The resultant spectra, plotted in Fig. \ref{fig:sed}a, show much flatter power-law indices (see Table \ref{table:sed}), with typical propagated uncertainties of $\sim$0.20 to 0.34 on the spectral index due to the extinction. The SEDs obtained are fairly smooth, with the $V$ band fainter than $i^{\prime}$ and $r^{\prime}$ throughout all outbursts. Although the SEDs have only three optical bands, the shapes are typical of the outer regions of an X-ray irradiated accretion disk \citep[e.g.][]{hynes2005}. We do not find any significant changes between the SEDs of the three outbursts.

\subsubsection{Optical/IR SED}

To construct the optical/infrared SEDs (see Fig. \ref{fig:sed}b), we first incorporate the GROND observations from \cite{grond} during the 2016 outburst (MJD 57454.3889). These data provide $g^{\prime}r^{\prime}i^{\prime}z^{\prime}JHK$ magnitudes obtained with a seeing of 1.3 arcsec, encompassing both the target and the blend stars BS1 and BS2. 

A simple blackbody fit to the GROND data indicates that no single temperature provides a fully satisfactory fit. A blackbody with a temperature of approximately 9000 K reproduces most of the optical part reasonably well; however, the infrared flux shows a slight excess relative to the blackbody prediction, while the optical $g^{\prime}$-band flux is somewhat underpredicted. Since BS1 has a similar infrared brightness to our target in quiescence but is fainter in the optical (Gaia non-detections, $G > 21$ mag), we investigated whether the NIR excess observed in the GROND data could arise from contamination by the blend stars.

The GROND observations were taken in 1.3$\arcsec$ seeing \citep{grond}, comparable to the separations of BS1 ($\sim$0.4$\arcsec$) and BS2 ($\sim$1$\arcsec$), so both sources necessarily contribute to some degree to the aperture photometry. BS1 has $K_S = 17.19 \pm 0.04$~mag in quiescence \citep{Torres2011}, whereas the GROND outburst flux corresponds to $K_S = 15.2 \pm 0.1$~mag, making BS1 $\sim$2 mag fainter than the observed blended value. This implies that BS1 contributes at the level of $\sim$16\% to the total $K_S$-band flux, corresponding to a brightening of $\sim$0.2 mag. Since the inferred NIR excess from our SED fit is $\sim$0.5 mag (Fig.~\ref{fig:sed}), BS1 alone can account for a substantial fraction of the excess but cannot explain it in full. BS2 is brighter than BS1 in the NIR \citep{Chaty2011} and may also contribute significantly to the GROND flux, although a larger fraction of its PSF will fall outside the photometric aperture due to its greater separation from the XRB. Without detailed information on the seeing and exact photometric setup of the published measurements, its precise contribution cannot be reliably quantified. 
Taken together, the observed NIR excess can be interpreted as a combination of unresolved contributions from BS1 and BS2, and possibly an additional intrinsic near-infrared component associated with the XRB itself. A plausible intrinsic origin for the remaining excess is optically thin synchrotron emission from a compact jet, which commonly contributes in the NIR during outburst \citep{Jain2001,routnir,2019ApJ...887...21S}, although irradiated disc emission may also play a role.

We also built SEDs using WISE data from \cite{john}, who subtracted non-outburst images to isolate the flux of the target in three epochs, one in 2016 and two in 2022, ensuring that there is no contamination from the nearby sources. For the three WISE detections reported in the literature, simultaneous optical observations are not always available. When no data are available within 24 hours, we use the closest available optical data points within 4 days of the WISE observation. We find that the infrared tail follows the trend observed in the $JHK$ bands from GROND, suggesting a consistent NIR extension. In addition to the outburst-phase SEDs, we construct a quiescent optical/NIR SED using the $K_s$-band measurement from \cite{Torres2011}, obtained on 2008 June 23 (MJD 54640), with the target deblended from BS1. To complement this, we incorporate the fitted (mean) quiescent magnitudes from our LCO observations in the $i^{\prime}$ and $r^{\prime}$-bands (Sec.~\ref{sec:dan}), finding that the overall SED shape is similar to that in outburst but at a much lower flux level. Finally, we include VISIR J8.9 and PAH2$\_$2 upper limits, along with our quasi-simultaneous $i^{\prime}$ and $r^{\prime}$-band detections from LCO, to further constrain the infrared emission. These SEDs summarise the optical/infrared characteristics of the source during both outburst and quiescence, incorporating the available data and accounting for possible contamination. 


\begin{figure}
    \centering
    \includegraphics[width = 1.0\columnwidth]{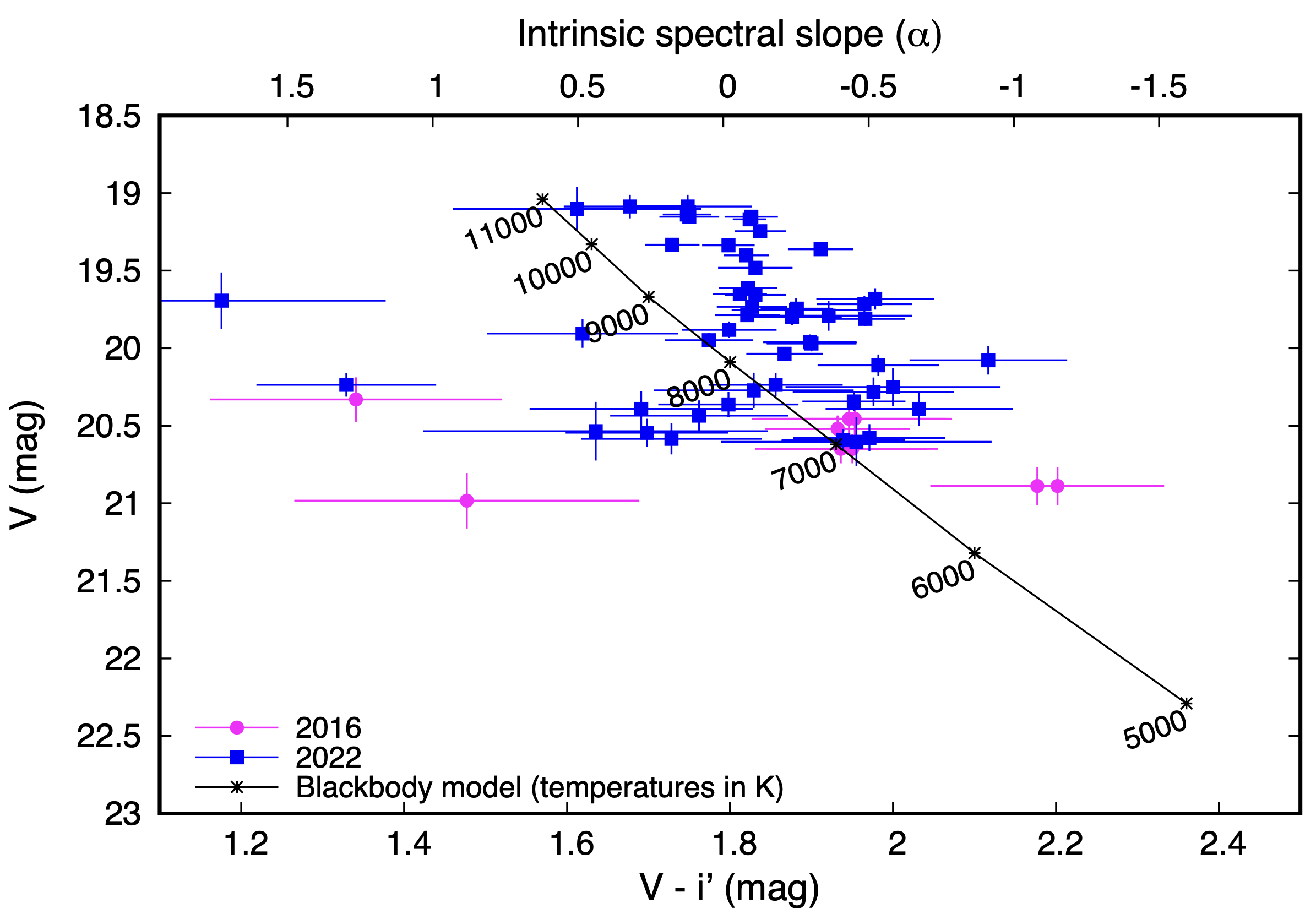}
    \caption{Color-magnitude diagrams (CMD) of IGR~J17091 covering the 2016 (magenta circles) and 2022 (blue squares) outbursts. Observed magnitudes are plotted, while the top axis spectral index is shown assuming an extinction value of $A_{V}$ = 5.36. The black solid line represents a blackbody model with different temperatures, which approximates the emission of an X-ray irradiated outer accretion disk.}
    \label{fig:cmd}
\end{figure}


    \label{fig:ox}

\subsection{Color-magnitude diagram} \label{sec:cmd}


We also examine the observed color-magnitude diagram (CMD) of IGR~J17091 to explore its color evolution during various outbursts. We use quasi-simultaneous, non-dereddened, $V$ and $i^{\prime}$-band data obtained within 24 hours of each other (see Fig. \ref{fig:cmd}). We account for optical extinction in the spectral indices using the most recently measured value of $A_{V} \sim$ 5.36 mag \citep[][]{wang2}. 

We overlay a model of a blackbody that is spatially uniform in temperature, with a constant area and a varying temperature, on the data \citep[see][]{Maitra2008, Russell11}. The slope of the model and the temperature range are determined solely by the extinction. However, the normalization of the model (up and down) depends on several uncertain system parameters, including the radius of the accretion disk, the distance to the source, the orbital period and angle of inclination, the filling factor of the disk and the warping of the disk. Due to these uncertainties, instead of fitting the data with the model, we choose a normalization that best aligns with the bottom envelope of the data \citep[see methods in][]{Zhang2019, baglio2020, cen, saikia1716, saikia1910apj, rout2025}. We find that the model approximates the slope of the data well, suggesting that the optical emission during outbursts in IGR~J17091 can largely be explained by a single-temperature blackbody, likely originating from the irradiated accretion disk. 
The temperature of the blackbody model corresponding to the optical color is mostly in the range of 7000--11,000 K, which encompasses the temperature required for hydrogen in the accretion disk to be ionized \citep{Lasota2001, Hameurynew}. 

\subsection{Optical / X-ray correlation} \label{sec:optX}

We use quasi-simultaneous multi-wavelength correlations of the source to further investigate the various components contributing to its optical emission during an outburst. For this analysis, we plot Swift/XRT ($2-10$ keV) 
X-ray count rates obtained within 24 hours of an LCO $r^{\prime}$-band magnitude measurement. All the X-ray count rates were converted to fluxes with the WebPIMMS tool \footnote{https://heasarc.gsfc. nasa.gov/cgi- bin/Tools/w3pimms/w3pimms.pl}, using a photon index of $\Gamma\sim$2.00 (as complete spectral state information is unavailable for all data points, and the majority of X-ray observations were acquired during the soft state), and a hydrogen column density value of $N_{\mbox{\scriptsize H}}=(1.537\pm0.002)\times 10^{22}\, \rm cm^{-2}$ \citep{wang2}. We note that since we have used the same energy range to convert the count rates to fluxes, a change in the photon index (e.g., from $\Gamma\sim$2.00 to $\Gamma\sim$1.60) does not alter the fluxes, and hence yields the same correlation coefficients (as long as the same $\Gamma$ is applied consistently across all count rates). Ideally, we would use a specific $\Gamma$ for each observation date, but for IGR J17091, we lack sufficient information to precisely determine its spectral state at different times during each outburst. In this process, any information on spectral evolution is lost; however, our primary focus is on the overall X-ray flux and the correlation coefficient.

\begin{figure}
    \centering
    \includegraphics[width = \columnwidth]{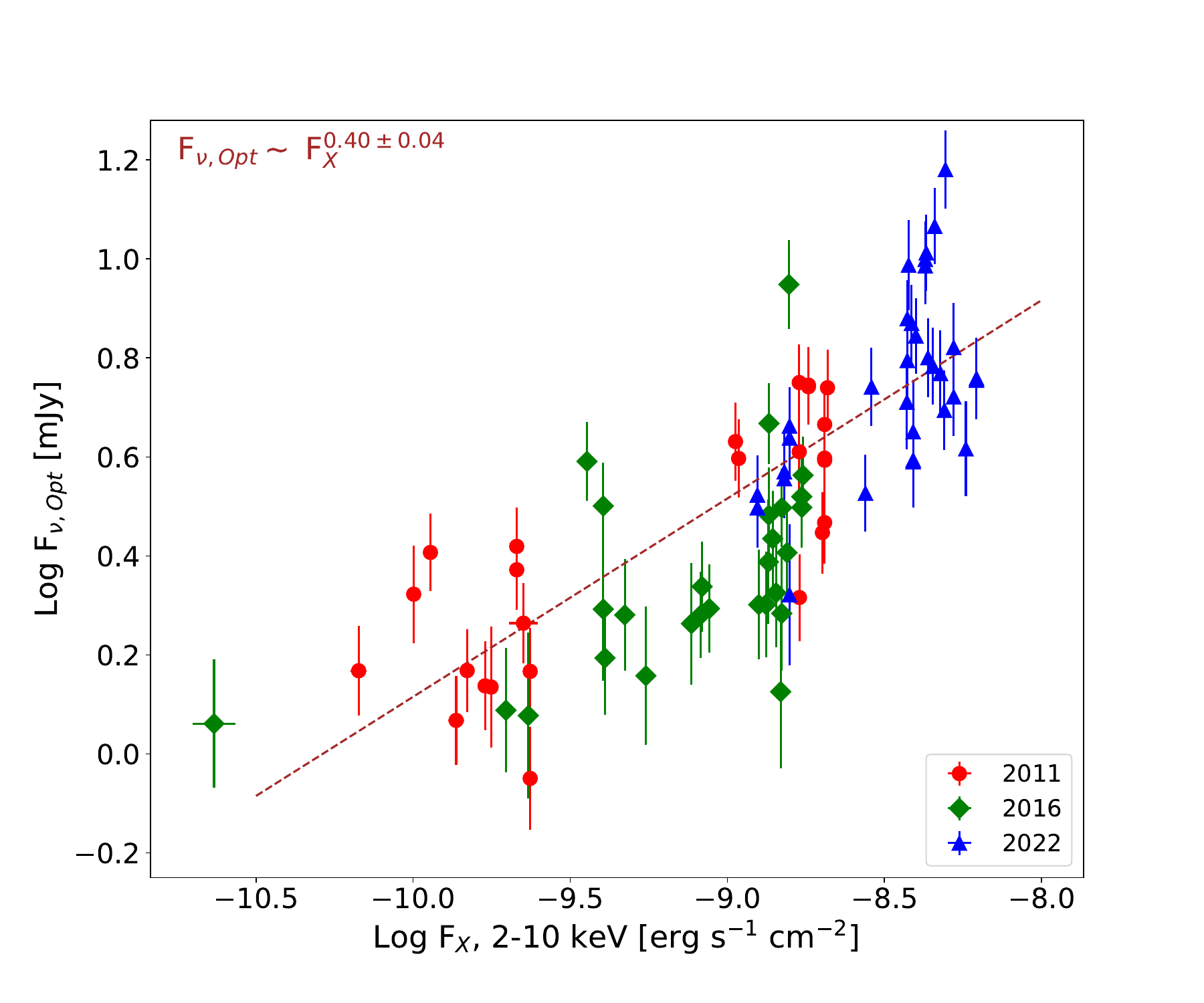}
    \caption{Optical/X-Ray correlation using LCO $r^{\prime}$-band and quasi-simultaneous Swift/XRT 2-10 keV flux for the three outbursts - 2011 (red circles), 2016 (green diamonds) and 2022 (blue triangles). The red dashed line represents the best-fit relation for the complete data; the corresponding equation is displayed at the top of the figure.}
    \label{fig:ox}
\end{figure}

\begin{table*}
\centering
\caption{Variability amplitude computed as the fractional root mean square (rms) in the flux. The method of \protect\cite{vaughan} was used for the high-cadence optical data of IGR J17091, and the quasi-simultaneous X-ray data binned at 1s (to trace the sub-second variability) and 120s (to match with the time-resolution of the optical high cadence data).}
\begin{tabular}{ l l l l l l l}
\hline
\hline
Date & Optical Time range   &  Data &  rms (Opt) & X-ray Time range &  rms (1s) &  rms (120s) \\
(MJD) & (UTC)  &  (N) &  (\%) & (UTC)  &  (\%) &  (\%) \\
\hline
\hline
$i^{\prime}$-band data  & & & &   & & \\
\hline
2022-04-02 (59671)	&  08:50:49.15 - 09:11:03.73	 & 	 15 & $*$ & 09:35:14.18 - 09:46:09.18 & 13.96$\pm$0.15 & 3.87$\pm$0.22\\
2022-04-03 (59672)	&  22:45:56.85 - 23:06:08.99	 & 	16 & 3.80$\pm$2.70 & 00:07:52.25 - 00:11:25.25$!$ & 13.68$\pm$0.26 & 3.05$\pm$0.10\\
2022-04-06 (59675)	&  08:55:56.42 - 09:16:09.67	 & 	15 & $*$ & 08:23:37.18 - 08:29:41.18 & 12.95$\pm$0.21 & 0.86$\pm$0.33\\
2022-04-08 (59677)	&  09:41:01.36 - 10:01:14.68	 & 	14 & 1.38$\pm$1.29 & 09:58:40.18 - 10:03:47.18 & 16.43$\pm$0.23 & 5.05$\pm$0.26\\
2022-04-11 (59680)	&  14:12:34.15 - 15:27:22.29	 & 	47 & 2.77$\pm$0.44 & 14:44:57.18 - 14:47:30.18 & 12.75$\pm$0.39 & 2.89$\pm$0.50 \\
2022-04-15 (59684)	&  07:34:42.53 - 07:54:53.25	 & 	15 & $*$ & 07:04:33.18 - 07:18:14.18 & 3.87$\pm$0.16 & 1.18$\pm$0.13 \\
2022-04-17 (59686)	&  15:03:47.69 - 15:53:24.58	 & 	45 & 5.67$\pm$0.78 & 13:20:53.18 - 13:29:06.18 & 2.51$\pm$0.24 & $*$\\
2022-05-07 (59706)	&  09:35:45.49 - 09:55:57.76	 & 	15 & 1.17$\pm$1.34 & $-$ & $-$& \\
2022-05-09 (59708)	&  05:29:17.00 - 05:49:27.98	 & 	15 & 8.24$\pm$0.65 & $-$ & $-$ & \\
2022-05-11 (59710)	&  08:33:53.48 - 08:54:06.88	 & 	12 & $*$ & $-$ & $-$ & \\
2022-05-15 (59714.2)	&  06:34:02.32 - 06:54:17.28	 & 	15 & $*$ & $-$ & $-$ &\\
2022-05-15 (59714.5)	&  13:35:42.58 - 13:57:39.24	 & 	30 & $*$ & $-$ & $-$ &\\
\hline
\hline
$r^{\prime}$-band data & & & & & &\\
\hline
2022-04-11 (59680)	&  14:45:53.28 - 15:21:06.62	 & 	 16 & $*$ & 14:44:57.18 - 14:47:30.18 & 12.75$\pm$0.39 & 2.89$\pm$0.50\\
2022-04-15 (59684)	&  07:56:24.66 - 08:16:37.29	 & 	 15 & $*$ & 07:04:33.18 - 07:18:14.18 & 3.87$\pm$0.16 & 1.18$\pm$0.13 \\
2022-04-17 (59686)	&  15:25:58.40 - 15:46:36.94	 & 	 15 & $*$ & 13:20:53.18 - 13:29:06.18 & 2.51$\pm$0.24 & $*$\\
2022-05-07 (59706)	&  09:57:33.58 - 10:17:51.73	 & 	15 & 2.94$\pm$1.55 & $-$ & $-$ &\\
2022-05-09 (59708)	&  05:50:58.45 -  06:11:07.267	 & 	15 & 2.23$\pm$2.00 & $-$ & $-$ &\\
2022-05-15 (59714)	&  06:55:47.83 - 07:16:02.44	 & 	13 & $*$ & $-$ & $-$ & \\
\hline
\end{tabular}
\par
\raggedright  $*$ The sample variance is smaller than the mean square error, so the calculated fractional rms flux deviation is consistent with being zero.\\
\raggedright $!$ The closest MJD falls on the next day (2022 April 4, MJD 59673).
\label{table:rms}
\end{table*}

\begin{figure*}
    \centering
    \includegraphics[width = 0.95\columnwidth]{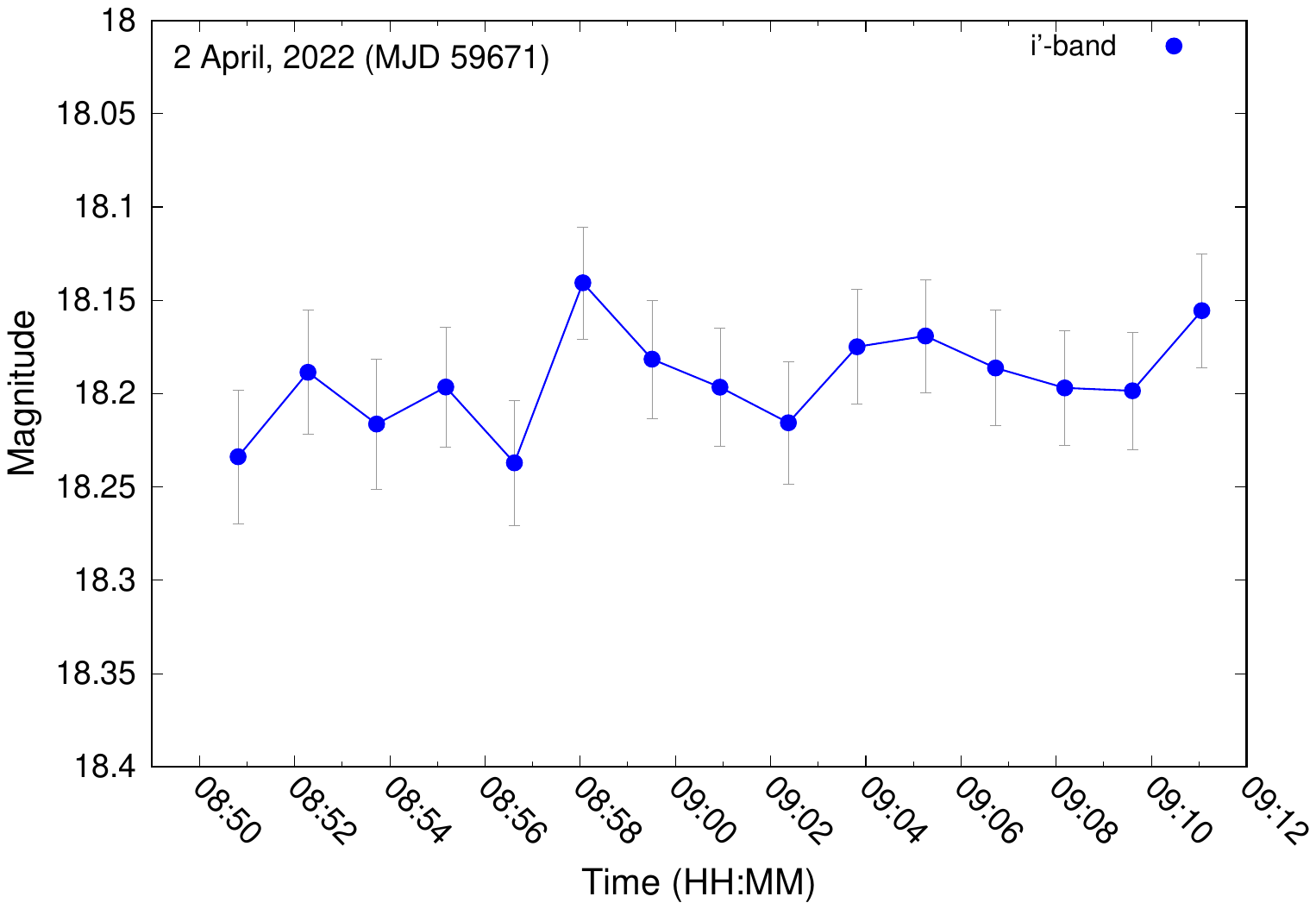}
    \includegraphics[width = 0.95\columnwidth]{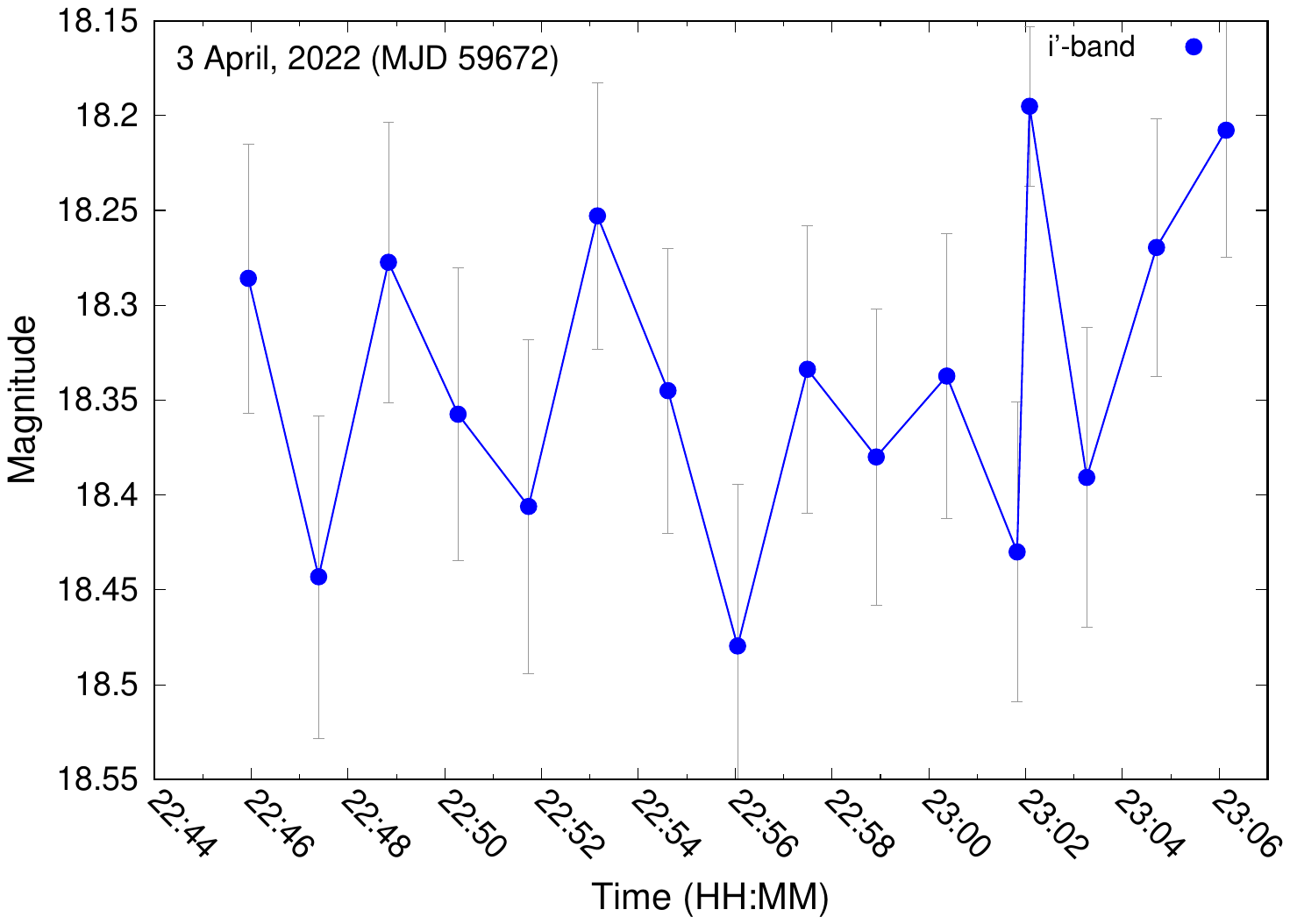}
    \includegraphics[width = 0.95\columnwidth]{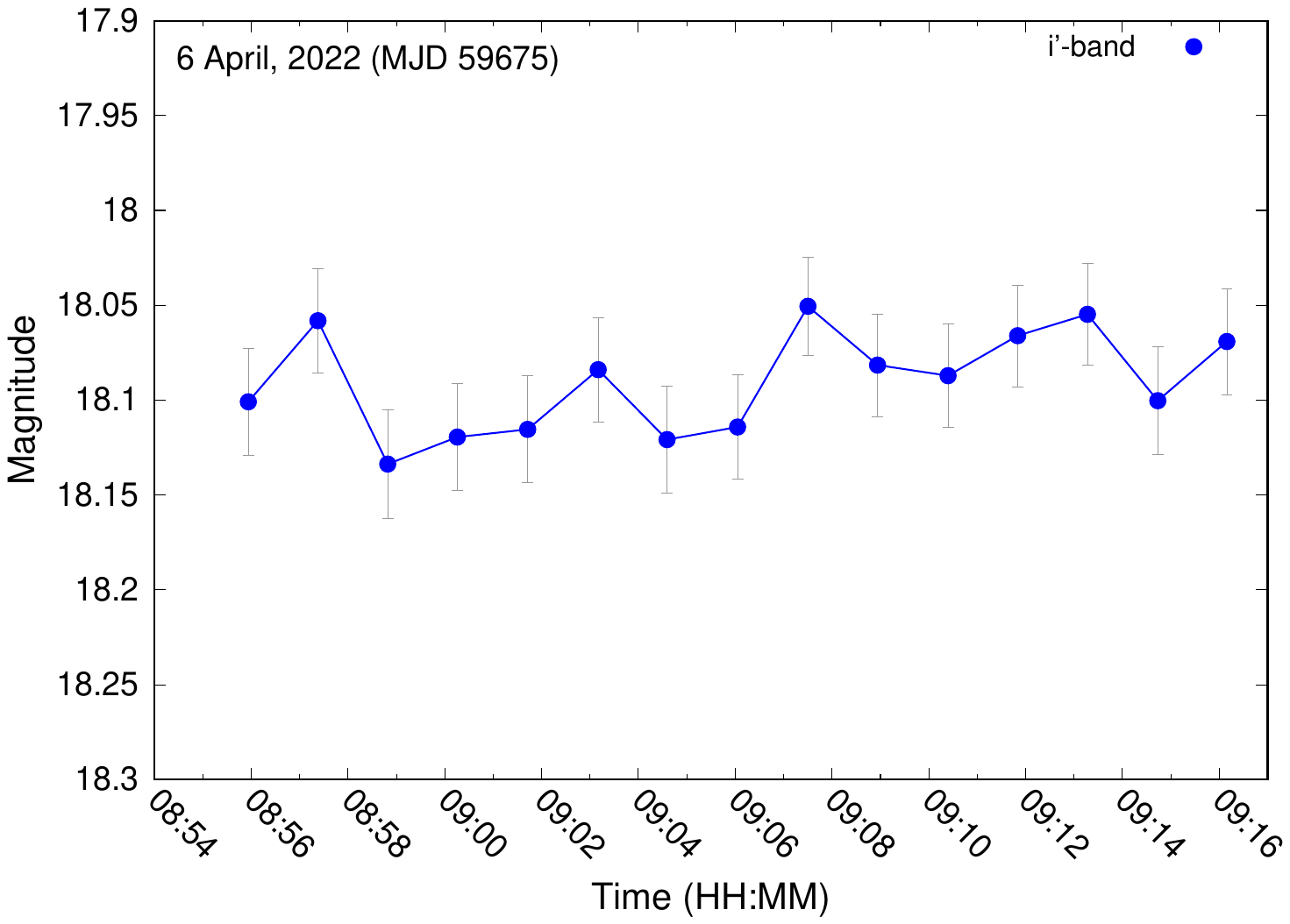}
    \includegraphics[width = 0.95\columnwidth]{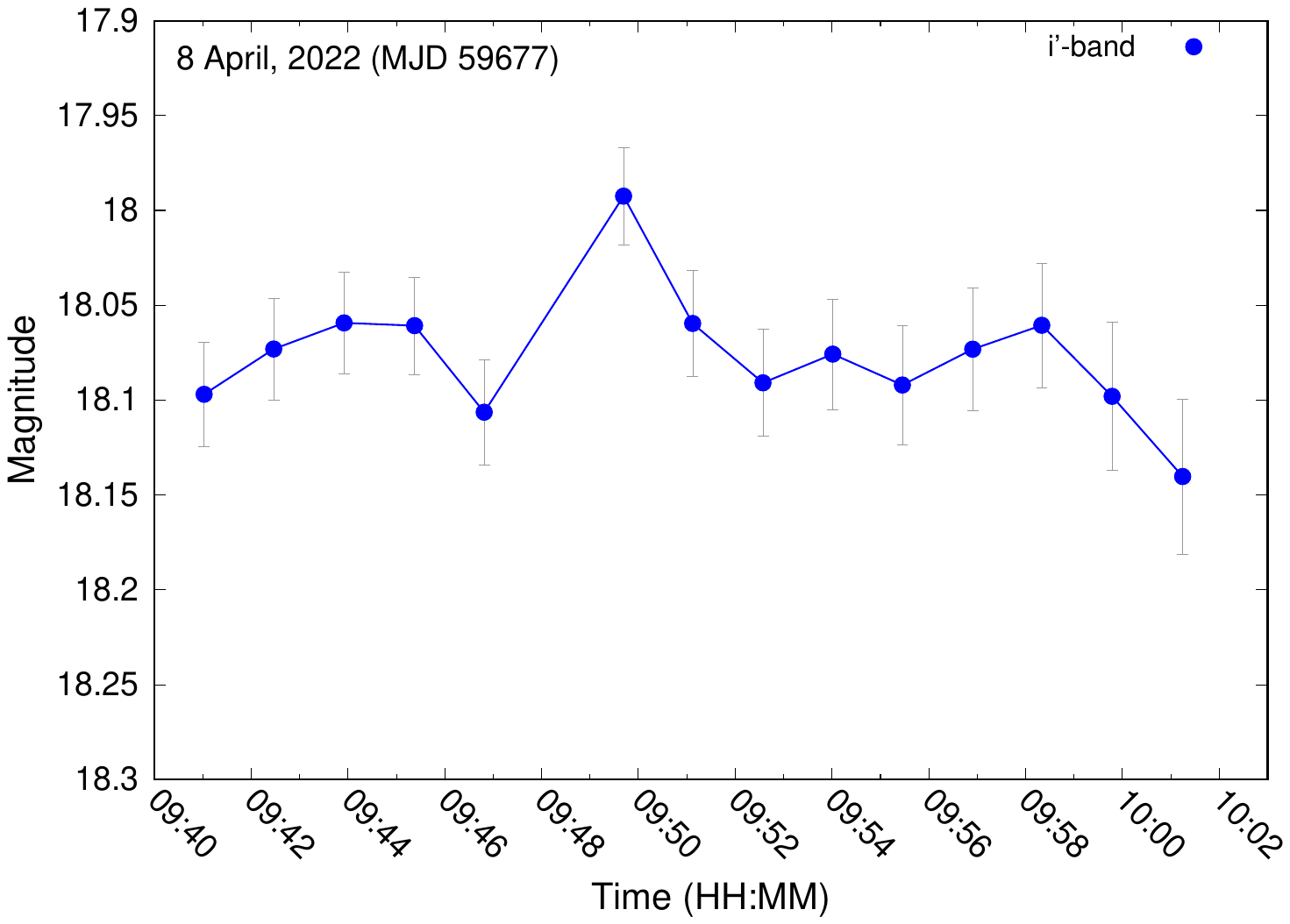}
    \includegraphics[width = 0.95\columnwidth]{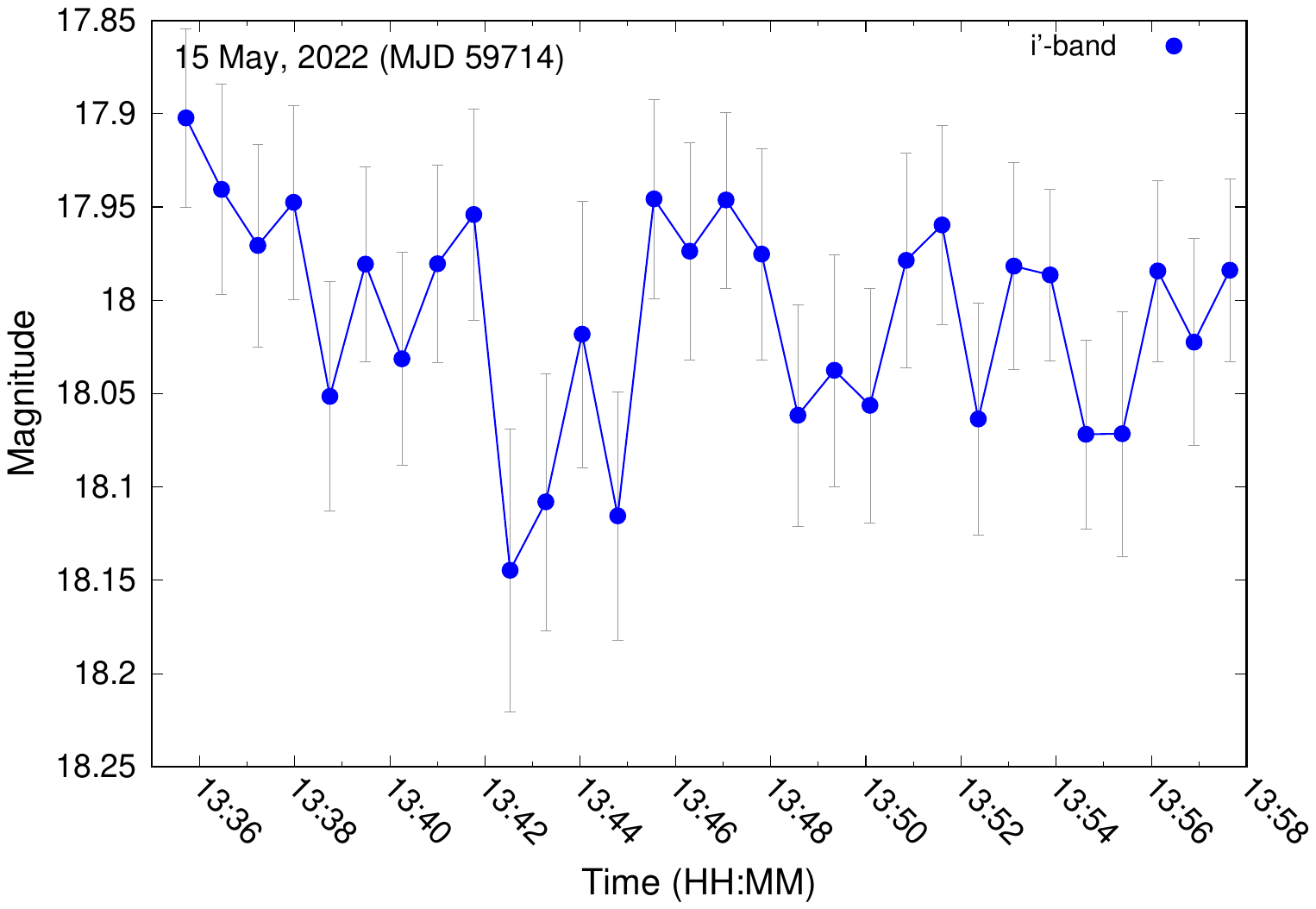}
    \caption{Optical high cadence observations in the $i^{\prime}$-band (blue circles) for all the days except MJD 59710 for which the error bars on individual points were quite large. The days when we had both $i^{\prime}$ and $r^{\prime}$ data are plotted in Fig. \ref{fig:optical_hc}. The Y-axes are consistently scaled to a range of 0.4 mag, and the X-axes are scaled to 23 minutes.
    }.
    \label{fig:optical_hc_i}
\end{figure*}

\begin{figure*}
    \centering
    \includegraphics[width = 0.95\columnwidth]{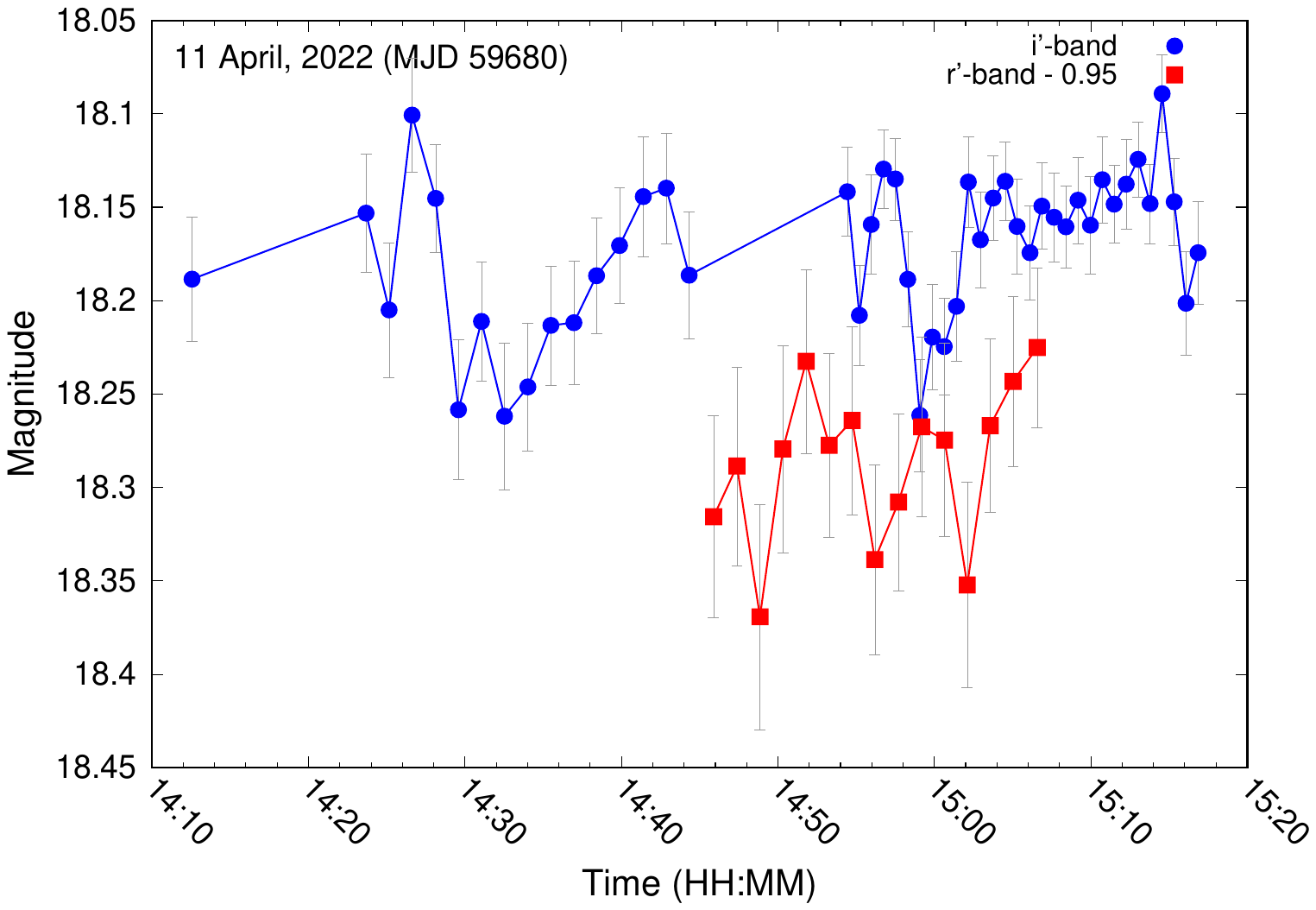}
    \includegraphics[width = 0.95\columnwidth]{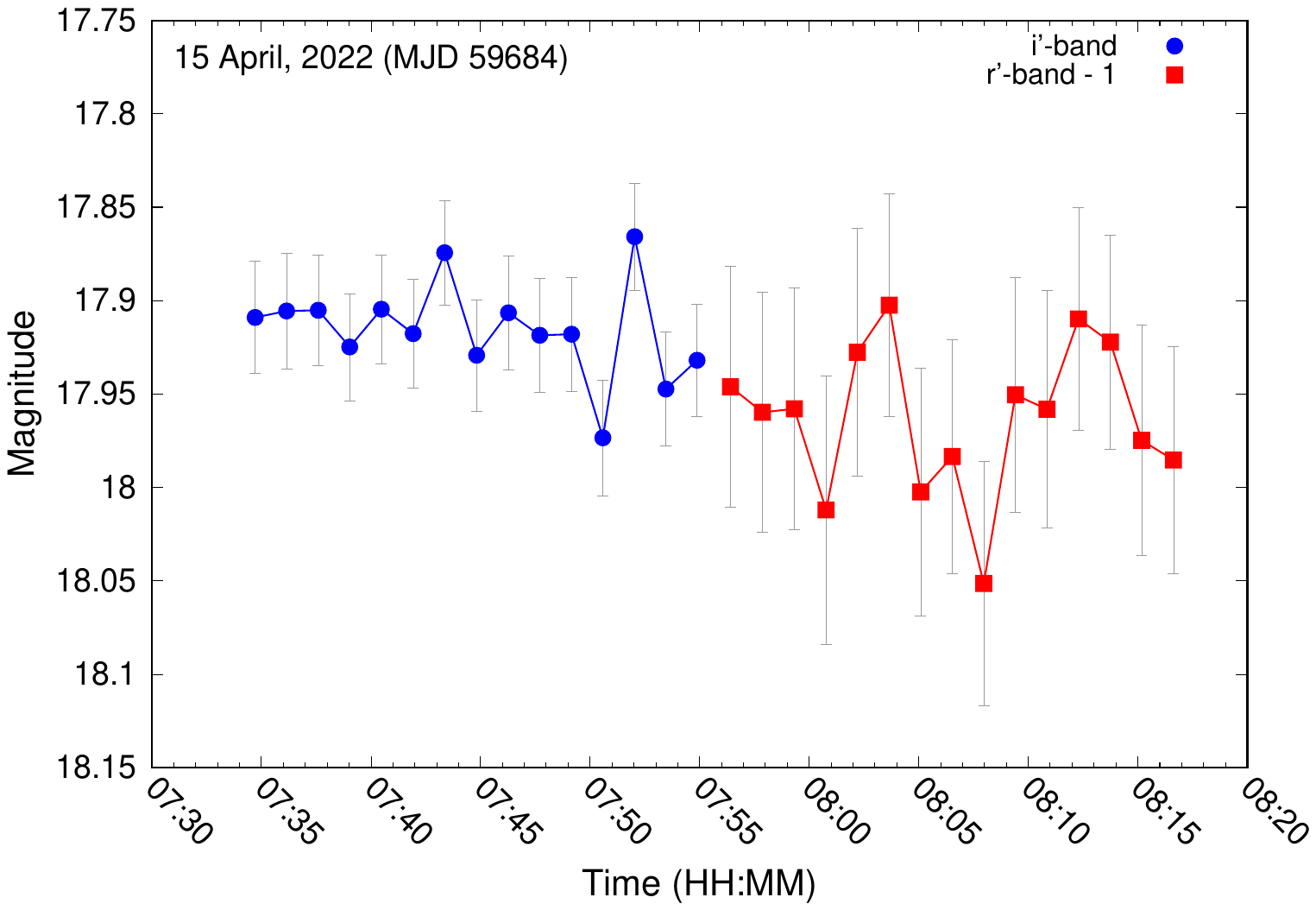}
    \includegraphics[width = 0.95\columnwidth]{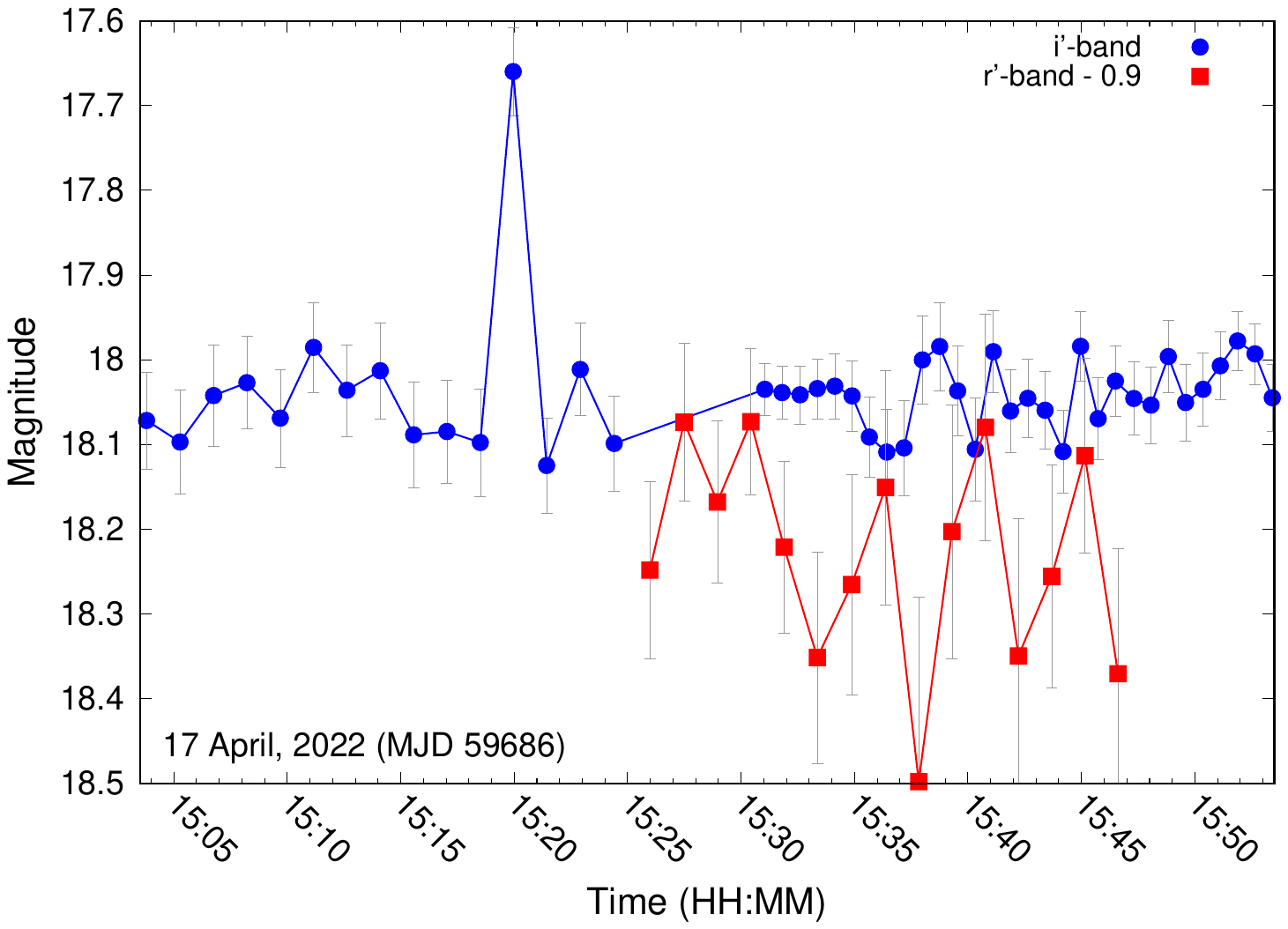}
    \includegraphics[width = 0.95\columnwidth]{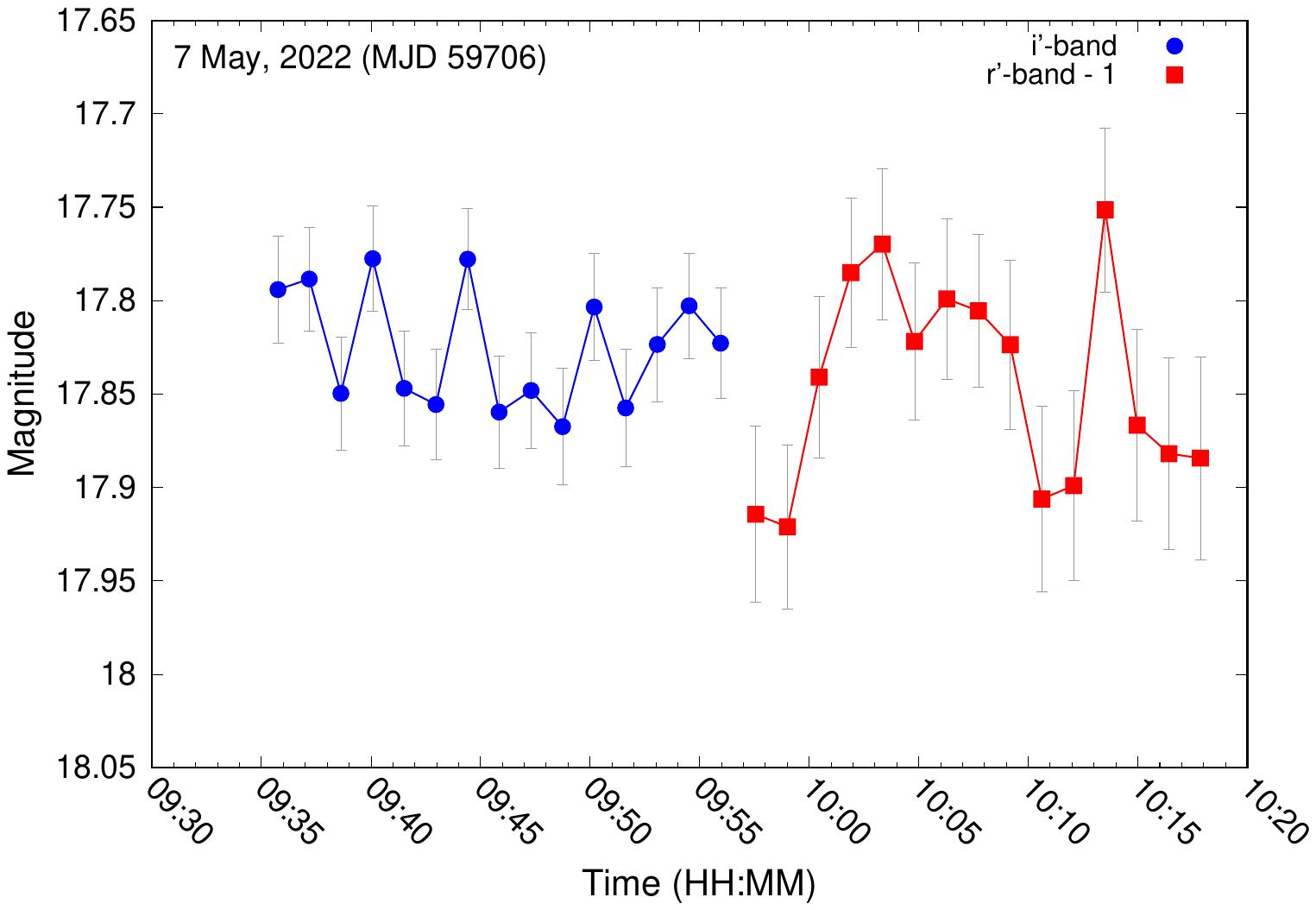}
    \includegraphics[width = 0.95\columnwidth]{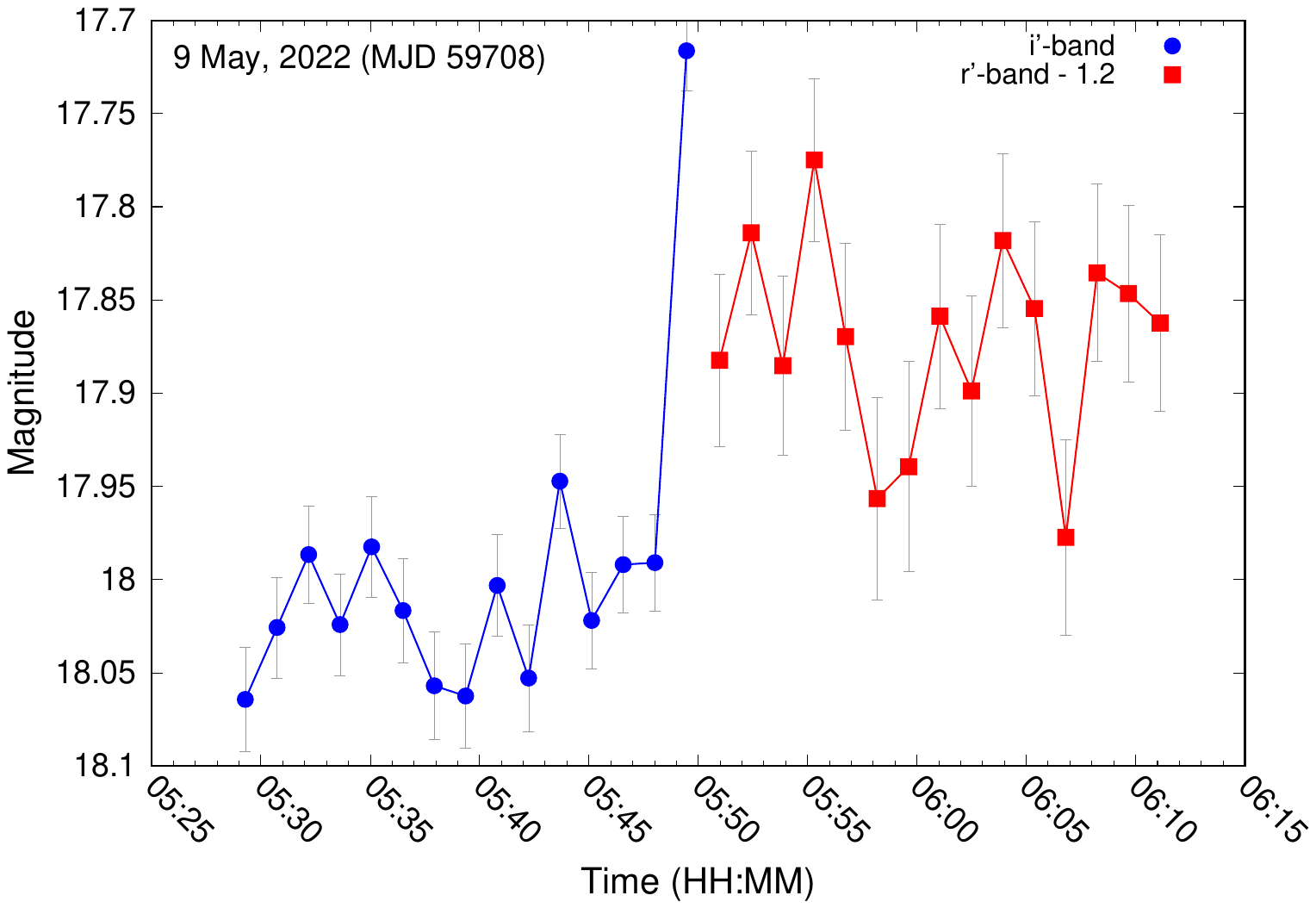}
    \includegraphics[width = 0.95\columnwidth]{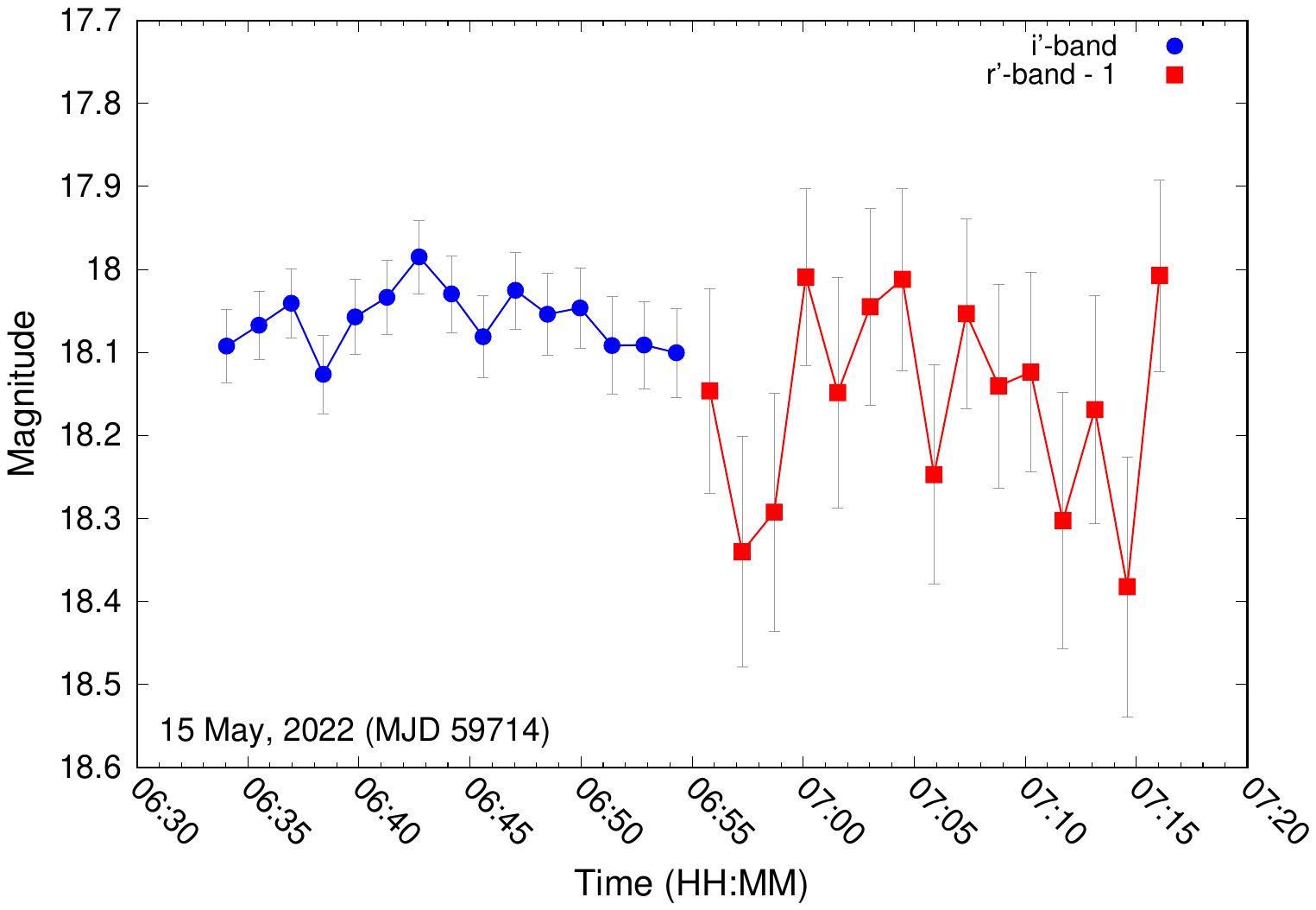}    
     
    \caption{Optical high cadence observations in the $i^{\prime}$-band (blue circles) and $r^{\prime}$-band (red squares). The X-axis is scaled to 50 minutes for each plot, except for MJD 59680, where it spans 1 hour 10 minutes. The Y-axis span is 0.4 mag for all plots except MJD 59686 and 59714, where a 0.9 mag span is used to capture the full dataset.}.
    \label{fig:optical_hc}
\end{figure*}




Despite the caveats, we observe a strong correlation between the X-ray and optical flux of IGRJ17091, with a Pearson correlation coefficient of 0.8 and a p-value of $10^{-10}$. All three outbursts individually follow very similar trends, yielding an overall power-law relation $F_{\mathrm{opt}} \propto F_{\mathrm{X}}^{\beta}$ (see Fig. \ref{fig:ox}). To account for the observed dispersion beyond the quoted measurement uncertainties, we fitted the correlation using a maximum-likelihood routine that includes an intrinsic scatter term added in quadrature to the optical observational errors. This method simultaneously optimizes the slope, intercept, and intrinsic scatter, providing a statistically consistent description of the data. The final best-fit slope is $\beta$ = 0.40 $\pm$ 0.04, with an estimated intrinsic scatter of $\sigma_{int}$ = 0.14 dex, corresponding to an additional $\sim$38\% variation in the optical flux at a given X-ray flux that cannot be explained by measurement uncertainties alone. This intrinsic scatter likely reflects short-term optical variability, differences in the emission regions contributing to the optical flux, or other unmodeled physical effects.

The measured slope suggests a disk origin for the optical emission in IGRJ17091 during outburst, although it remains unclear whether the emission arises primarily from a viscous accretion disk, an X-ray irradiated disk, or a combination of both (see Section\ref{sec:emission} for a detailed discussion).

For IGR J17091, we do not find a significant correlation between the hard X-ray flux (Swift/BAT, 15–50 keV) and the optical flux. This lack of correlation may be attributed to the highly variable nature of IGR J17091, where the use of daily-averaged Swift/BAT fluxes may not adequately capture the rapid variability of the source, potentially obscuring any underlying relationship.

\subsection{Multi-wavelength high-cadence data} \label{sec:hc}

In addition to the long-term monitoring campaign of IGR J17091, we conducted high-cadence photometry (minute-timescale  observations) of the source using the 2-m Faulkes Telescopes during the 2022 outburst. These observations were carried out on 11 days (MJD 59671, 59672, 59675, 59677, 59680, 59684, 59686,  59706, 59708, 59710, and 59714), primarily in the $i^{\prime}$-band (see Fig. \ref{fig:optical_hc_i} and \ref{fig:optical_hc}) and occasionally in the $r^{\prime}$-band (see Fig. \ref{fig:optical_hc}). On most of these days, we had $\sim$15 detections in $\sim$20 mins with a time resolution of $\sim$ 85 seconds (i.e. a frequency range of 0.001--0.01 Hz). The variability in all these days looks quite structured, however the cadence and signal-to-noise of these optical observations do not permit the construction of a meaningful power spectrum. As a result, we cannot quantitatively assess any periodicity in the variability or demonstrate that the observed fluctuations are oscillatory in nature.


\begin{figure}
    \centering
    \includegraphics[width=\columnwidth]{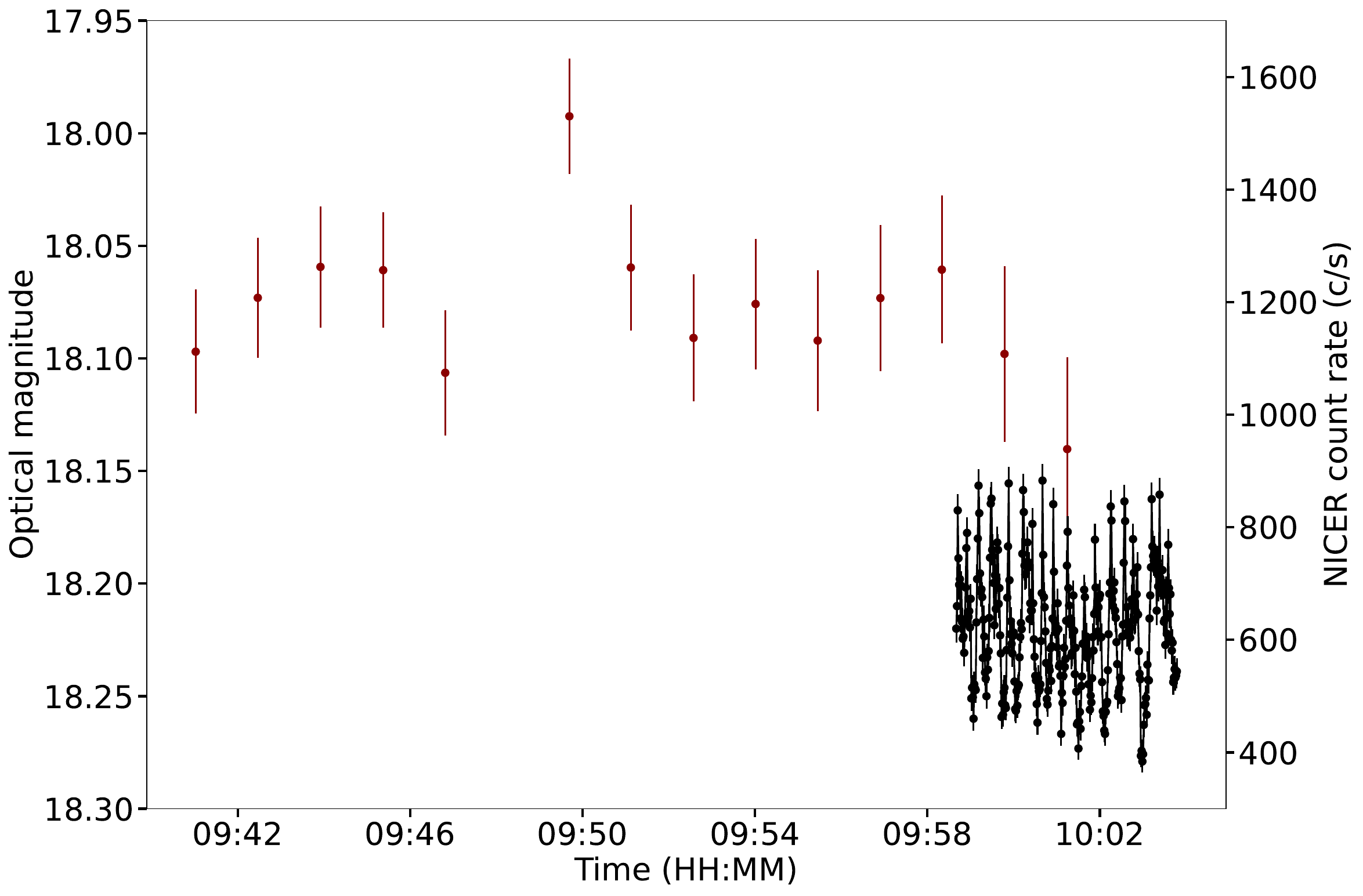}
    \includegraphics[width=\columnwidth]{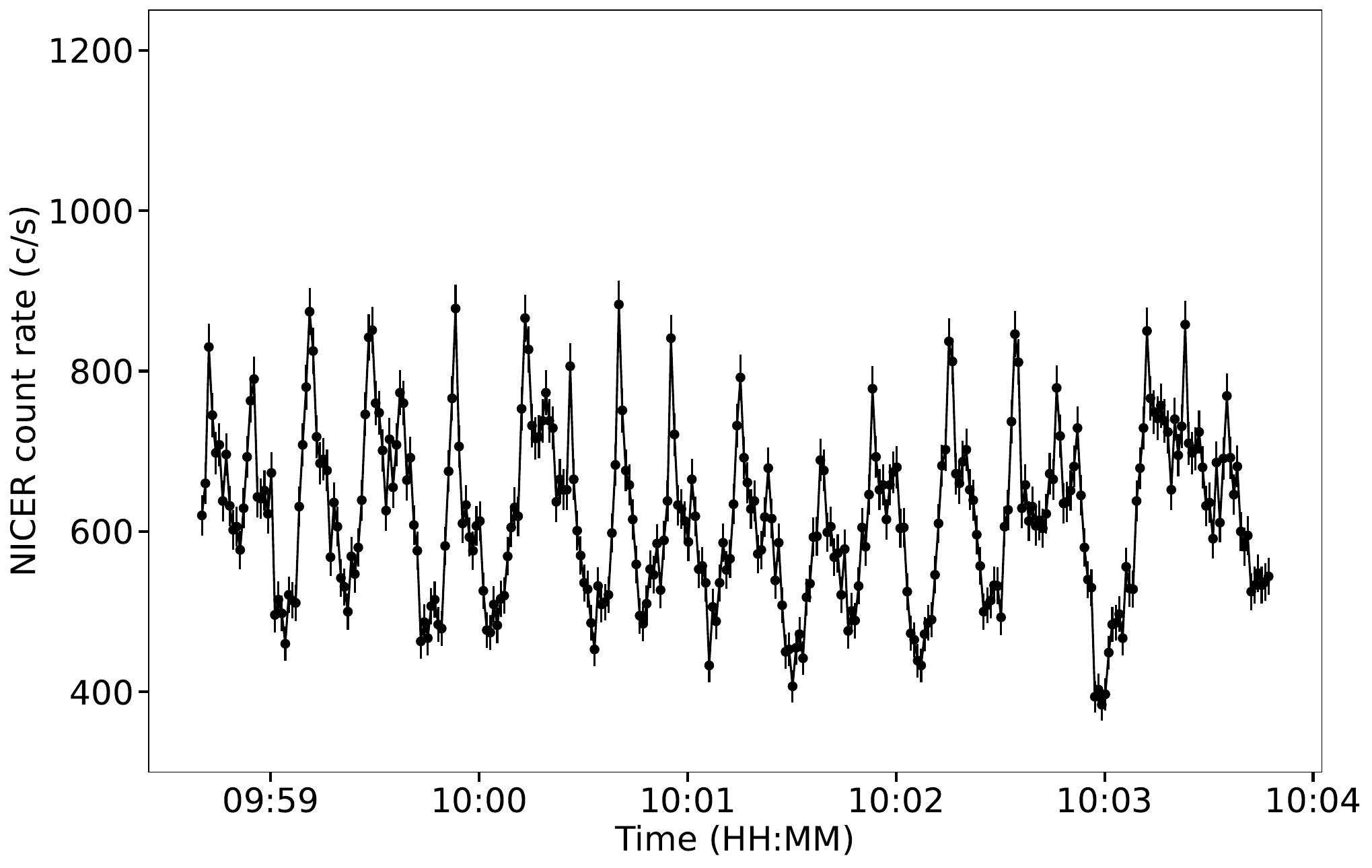}
    \caption{Top panel: Overlapping LCO and \textit{NICER} (1s binned; in black) data on MJD 59677. Bottom panel: Zoom of the \textit{NICER} light curve shown in the top panel (1s binned).}
    \label{fig:nicer_lc}
\end{figure}

\begin{figure}
    \centering
    \includegraphics[width = 0.9\columnwidth]{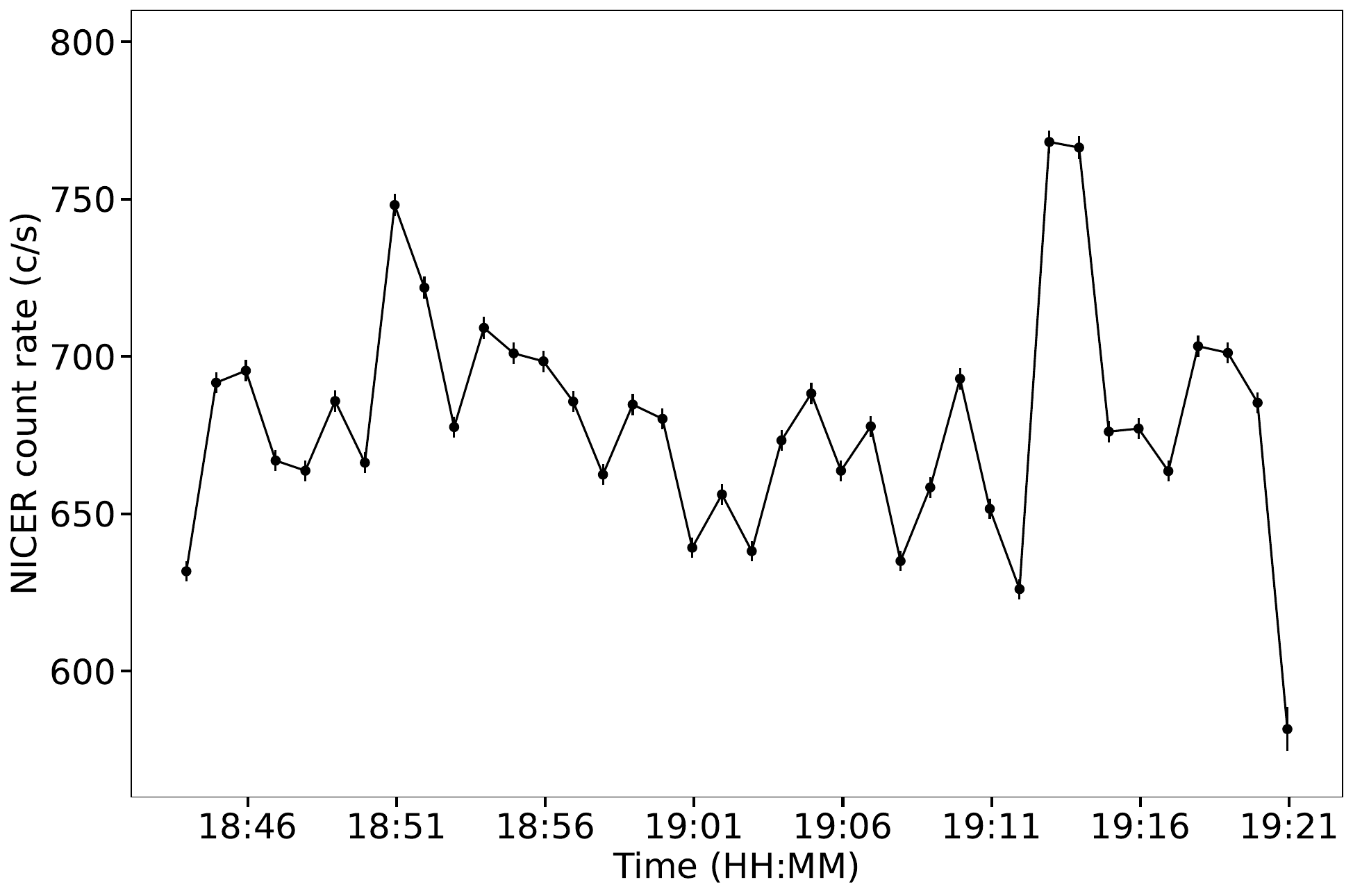}
    \caption{Representative quasi-simultaneous \textit{NICER} 120s binned (black points) light curve corresponding to the ObsID 5202630116 on MJD 59677.}
    \label{fig:nicer_lc2}
\end{figure}

We calculated the optical fractional rms deviation in the flux for these dates following the methods described by \cite{vaughan} and \cite{gandhi2010}. The resulting rms values (in the frequency range of 0.001--0.01 Hz) ranged from approximately 1 to 8 per cent, although we have a 3$\sigma$ detection of variability on only 3 dates (see Table \ref{table:rms}). To verify that the short-timescale variability observed from IGR J17091 during these epochs is not consistent with photometric noise, we analysed all neighbouring stars of similar brightness ($i^{\prime}$ magnitudes 17.5--18.5) within 100 arcsec of the source on two high-cadence nights (MJD 59680 and 59686). This yielded $\sim$125 suitable comparison stars per night. For each comparison star, we computed the optical fractional rms deviation in the flux over the same high-cadence intervals as for the target.
On MJD 59680, IGR J17091 showed rms of 2.77 $\pm$ 0.44 $\%$, while the comparison stars had a mean rms of 1.60 $\%$, with 23 stars returning zero intrinsic variance because their measured scatter was smaller than the photometric uncertainties (i.e. these stars are found to be non-variable at the sensitivity of the data set). On MJD 59686, again the source exhibited a high rms value of 5.67 $\pm$ 0.78 $\%$, compared to a mean neighbour value of 1.70 $\%$, with 71 zero-variance stars. In both epochs, the intrinsic variability of IGR J17091 exceeds that of the field population by more than a factor of $\sim$1.7, confirming that the observed variability is real and not completely driven by noise.



\begin{figure}
    \centering
    \includegraphics[width = \columnwidth]{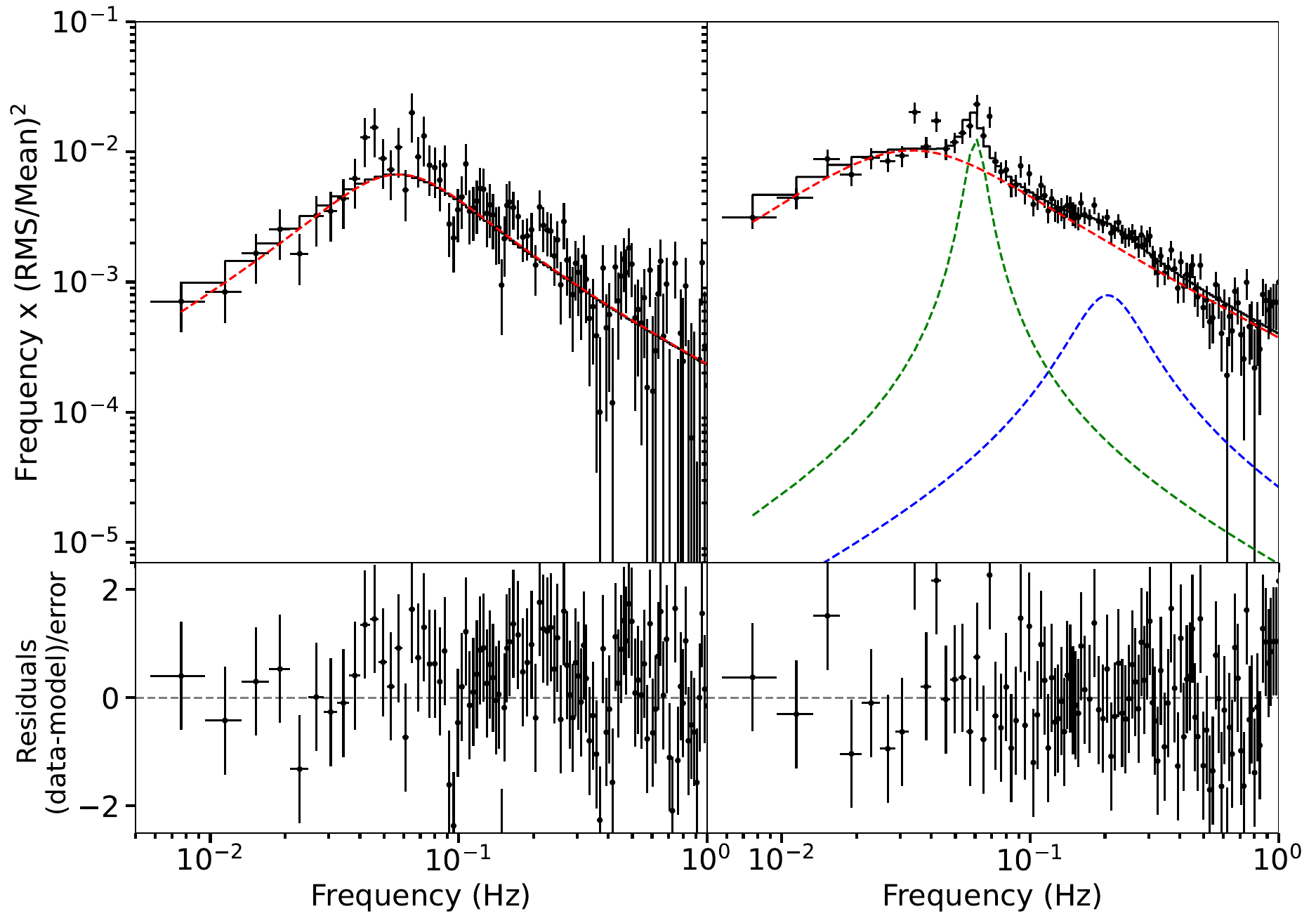}
    \caption{Two representative PDS of \textit{NICER} observations of IGR~J17091 during the dates we have quasi-simultaneous high-cadence LCO observations. The left panel (ObsID 5202630112) shows an example of a PDS with a single broadband Lorentzian with best-fit parameters $\nu_{0}=0.041\pm0.004$~Hz, $w=0.079\pm0.008$~Hz and $N=0.013\pm0.001$, and a $\chi^2/dof = 136.53/146$. The right panel (ObsID 5202630116) shows a PDS with a broadband component and two narrow peaks. The best-fit parameters are $\nu_{0}=0.0155\pm0.003$~Hz, $w=0.059\pm0.003$~Hz and $N=0.025\pm0.001$ (red component); $\nu_{0}=0.060\pm0.002$~Hz, $w=0.012\pm0.004$~Hz and $N=0.034\pm0.001$ (green component); $\nu_{0}=0.20\pm0.02$~Hz, $w=0.13\pm0.04$~Hz and $N=0.0007\pm0.0001$ (blue component); and a $\chi^2/dof = 161.83/140$. Dashed lines represent the best-fit Lorentzians.}
    \label{fig:nicer_pds}
\end{figure}

\subsubsection{Quasi-simultaneous X-ray variability}

We also compare the seven quasi-simultaneous \textit{NICER} light curves to the high-cadence optical observations with LCO. Unfortunately, there is only one slightly overlapping LCO and \textit{NICER} light curve (\textit{NICER} ObsID 5202630116, MJD 59677, shown in the top panel of Fig. \ref{fig:nicer_lc}). We find that in all the seven \textit{NICER} observations, the same variability pattern is observed on time-scales of seconds (bottom panel of Fig. \ref{fig:nicer_lc}). This pattern is consistent with the so-called \textit{Class V} variability presented in \cite{court2017} and \cite{wang2}. \textit{Class V} variability is characterized by the presence of two types of flares, one short, strong, and single-peaked flare, and another longer, more complex double-peaked flares lasting tens of seconds, with peaks separated by $\sim$12 mHz. The peak-to-peak time we found in this pattern is of a few tens of seconds, consistent with that presented in \cite{court2017}. The measured fractional rms deviation in the X-ray flux at 1s resolution is very high, lying in the range of 2.5 to 16.4 per cent.

Since the optical variability is studied on time-scales of minutes, we also study the 2-min binned \textit{NICER} light curves. We find a different variability pattern in all seven observations (see Fig. \ref{fig:nicer_lc2} for a representative light curve). These light curves show strong peaks with amplitudes of $\sim$60$-$100 c/s followed by several small peaks with amplitudes of a few tens of c/s. The time between strong peaks is somewhat greater than thousands of seconds, while the time between short peaks is a few hundred seconds, but it varies due to the irregular shape of the peaks.

The fractional rms deviation in the X-ray flux at such timescales is in the range of 0.8 to 5 per cent, comparable with the rms measured for the optical high cadence data. The highest optical fractional rms of $8.24 \pm 0.65$ per cent, measured on May 9, appears to be dominated by a single bright data point in the light curve, as does the next highest value of 5.67 per cent. In contrast, the April 11 observation does not exhibit such outliers and shows a significant rms of $2.77 \pm 0.44$ per cent, consistent within errors with the simultaneous X-ray rms. This is the only instance where optical and X-ray rms values are consistent within uncertainties; on other dates, the optical rms is either lower or poorly constrained.

\subsubsection{Power density spectra}

We finally study the PDS of the seven \textit{NICER} observations. To fit the power spectra, we use a multi-Lorentzian function. We use the characteristic frequency of the Lorentzians defined in \citet{Belloni02a}, $\nu_{\mbox{\scriptsize max}}=\sqrt{\nu_{0}^2 + (w/2)^2} = \nu_{0} \sqrt{1+1/4Q^{2}}$, where $w$ represents the full-width at half-maximum of the Lorentzian, $\nu_{0}$ the centroid frequency of the Lorentzian, and the quality factor $Q$ is defined as $Q = \nu_{0}/w$. The significance of the different Lorentzians are determined by the Normalization parameter, $N$. We consider a component significant if the S/N ratio of $N$ is higher than 3. Two observations (ObsIDs 5202630123 and 5202630125) showed little or no significant broadband variability ($<$3$\sigma$ in the 0.01$-$64.0 Hz frequency band), while the other five observations (ObsIDs 5202630112, 4618020801, 5202630114, 5202630116 and 5202630119) are characterised by a broad component (red Lorentzian in Fig. \ref{fig:nicer_pds}). A narrow component (green Lorentzian in Fig. \ref{fig:nicer_pds}) peaking between 0.06$–$0.10 Hz with a fractional rms of $\sim$5–12\% was also detected in three observations (ObsIDs 4618020801, 5202630114, and 5202630116), as also observed in Fig. \ref{fig:nicer_lc}b, consistent with a feature previously reported by \citet{wang2}. In addition, we also found a broad Lorentzian peaking at $\sim$0.2 Hz (ObsIDs 5202630116).

\section{Discussion} \label{sec:dis}

In this work, we present for the first time a comprehensive study of the optical emission from IGR J17091, covering three of its outbursts to investigate the dominant origin of its optical emission. Notably, this is the first-ever detailed optical study of any heartbeating BHXB, as the only other known example, GRS 1915+105, is heavily obscured by extinction, making optical detections out of reach with current facilities. We detect evidence of quasi-periodic optical modulations in the available high cadence (minute-timescale) data. This could potentially be linked to reprocessed and smeared X-ray variability, suggesting a connection between the exotic, highly structured fast X-ray variability and the optical emission in these heart-beating systems. Confirming this scenario requires higher time-resolution optical data. In the following sub-sections, we discuss the implications of these findings for the origin of the optical emission, constraints on the extinction towards IGR J17091, insights into its distance, and the nature of its short-term optical variability.

\subsection{Optical emission mechanism} \label{sec:emission}

In BHXBs during outburst, optical emission can arise from multiple sources, including X-ray reprocessing by the outer accretion disk \citep[e.g.][]{Cunningham1976, vrtilek}, thermal emission from a viscous accretion disk \citep[e.g.][]{ss,Accretion_Power}, synchrotron emission from a compact jet in the HS \citep[e.g.][]{markoff01,Russell2006,Kalemci2013,saikia2019} and sometimes in transitional states \citep[e.g.][]{fender04,Russell2020}, a hot inner flow \citep[e.g.][]{veledina2013}, and the companion star during quiescence \citep[e.g.][]{casares}. To disentangle these contributions, we employ various diagnostic methods, including CMDs, SEDs, and optical/X-ray correlations.

We examine the CMD of IGR J17091 and compare it with a blackbody model representing an accretion disk. This model describes the evolution of a constant-area blackbody, spatially uniform in temperature, as it undergoes heating and cooling. 
Fig. \ref{fig:cmd} reveals that the slope of the model closely matches the observed trend, following a pattern consistent with thermal blackbody radiation. We interpret this blackbody emission as originating from the outer accretion disk. The inferred temperatures range mostly between 7000 and 10,000 K, sufficient to fully ionize hydrogen in the disk, and to ensure the conditions necessary for the disk instability model \citep[DIM;][]{Lasota2001} to drive a complete outburst.

Examining the optical SEDs of IGR J17091, we find that the inferred spectral indices are highly sensitive to the choice of extinction towards the source (see also Section \ref{sec:extinction}). Typically, we expect a spectral index of $\sim$1 (with values ranging from $-$1 to +2 depending on the disk temperature) if optical emission comes from reprocessing in the X-ray irradiated outer accretion disk \citep[e.g.][]{hynes2005}, a slope in the range of 0.3 $< \alpha <$ 2.0 for a viscously heated outer disk \citep[e.g.][]{Accretion_Power}, and a negative slope $\alpha \sim$ $-0.7$ if the dominant emission is optically thin synchrotron from the jet \citep[e.g.][]{Gandhi11}. An intermediate slope is obtained when a combination of these processes contributes to the optical emission \citep[e.g.][]{saikia1716}. Due to the strong dependence of the slope on the chosen $N_{\mbox{\scriptsize H}}$ values, and the high uncertainity associated with it, the SEDs alone do not provide a clear distinction between different emission mechanisms in IGR J17091. However, using the most recent and updated $N_{\mbox{\scriptsize H}}$ value from \cite{wang2}, we find that the resulting spectral slopes are consistent with a disk origin for the optical emission.

We also analyze the quasi-simultaneous optical/X-ray correlation in IGR~J17091 and find a significant relationship between the Swift/XRT 2–10 keV flux and the optical flux. Generally, the power-law index of the optical/X-ray correlation can provide insights into the different emission mechanisms responsible for the overall flux of the system. If the optical emission is dominated by an X-ray irradiated accretion disk, simple geometrical assumptions predict $\beta \sim 0.5$ \citep{VanParadijs1994}. Conversely, if the optical emission is primarily from a viscously heated accretion disk, the power-law index is expected to be $\beta \sim 0.3$, with a dependence on wavelength \citep{Russell2006}. Newer theoretical studies indicate that the power-law index of the correlation can slightly vary depending on whether the optical emission originates from the Rayleigh–Jeans tail or closer to the peak of the blackbody, but it is generally expected to be $<0.7$ if the emission is dominated by the disk \citep[for a detailed discussion, see][]{Coriat2009, Shahbaz2015, be}. The correlation index is predicted to be $\beta \geq 0.7$, if the optical emission is mainly optically thin synchrotron radiation from a relativistic jet \citep{corbel2003, Russell2006}. For IGR J17091, we find that the correlation follows a power-law trend with a slope of $\beta = 0.40 \pm$ 0.04. This suggests that the dominant mechanism of the optical emission in IGR~J17091 could be either an X-ray irradiated accretion disk, a viscous disk, or a combination of both.

\subsection{Extinction towards IGR J17091} \label{sec:extinction}

The equivalent hydrogen column density towards IGR~J17091 is still uncertain. The $2^{\circ} \times 2^{\circ}$ weighted averaged Galactic absorption in the direction of the source is $N_{\mbox{\scriptsize H}} = 6.0 \times 10^{21} \rm cm^{-2}$ \citep{kalberla,rodri}. The initial X-ray observations of IGR~J17091 showed similar values, with $N_{\mbox{\scriptsize H}} \sim 7.0 \times 10^{21} \rm cm^{-2}$ \citep{corbelatel}. However, detailed spectral fitting of X-ray observations of the source in many independent studies later revealed that the value of $N_{\mbox{\scriptsize H}}$ could be much higher, consistent with a large distance to the source in the Galaxy. For example, while some studies have found the hydrogen column density to be $N_{\mbox{\scriptsize H}}=(1.1\pm0.3)\times 10^{22}\, \rm cm^{-2}$ \citep{capi,rodri,krimm}, a few others have reported even higher numbers such as $N_{\mbox{\scriptsize H}}=(0.97-1.7)\times 10^{22}\, \rm cm^{-2}$ \citep{iyer}, $N_{\mbox{\scriptsize H}}=(1.58\pm0.03)\times 10^{22}\, \rm cm^{-2}$ \citep{xu} and $N_{\mbox{\scriptsize H}}=(1.537\pm0.002)\times 10^{22}\, \rm cm^{-2}$ \citep{wang2}. 


\begin{figure}
    \centering
    \includegraphics[width = 1.0\columnwidth]{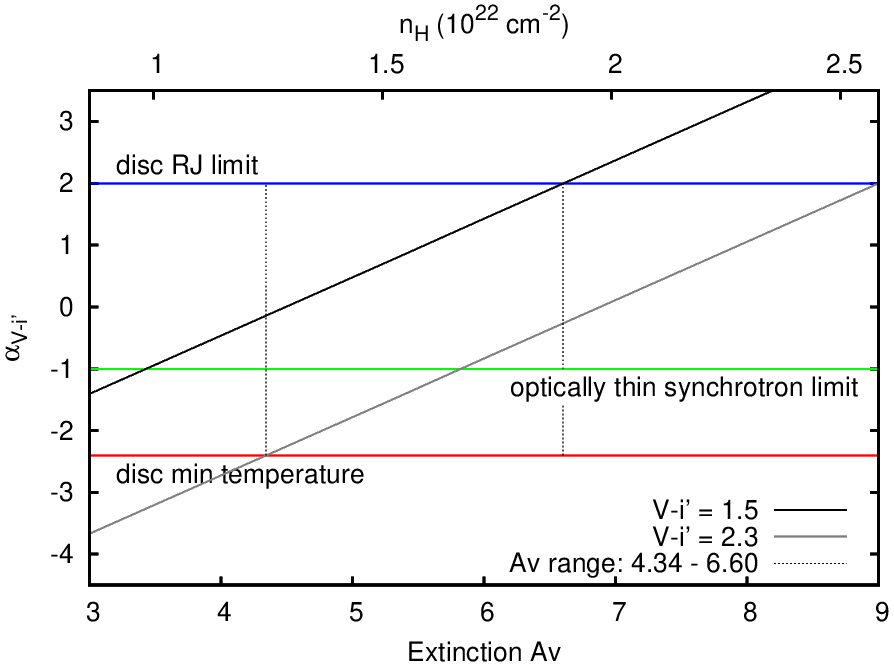}
    \caption{Slope of the optical SED vs extinction coefficient in $V$-band during the 2022 outburst. The three solid horizontal lines represent the expected boundaries for the optical spectral index - blue/top line (Rayleigh jeans limit of the accretion disk), green/middle line (optically thin synchrotron limit), and red/bottom line (minimum temperature possible for the disk). The two diagonal lines indicate the range of optical colors observed in IGR J17091, as derived from the CMD.}
    \label{fig:dave}
\end{figure}


Optical analysis can sometimes provide an independent and additional insight into extinction, particularly by examining the slope of the optical spectrum. For IGR J17091, we first constructed optical SEDs using the commonly adopted literature value of N$_{\mbox{\scriptsize H}}=(1.1\pm0.3)\times 10^{22}\, \rm cm^{-2}$ \citep{capi,rodri,krimm}. This resulted in highly negative spectral slopes ($\alpha_{V-i^{\prime}}$ ranging from $-$1.0 to $-$2.0), which are atypical for BHXBs in outburst, although we note that the propagated uncertainty on the spectral index is very large ($\sim$1.0), primarily because the adopted extinction value (A$_V$ = 3.83 $\pm$ 1.05) carries a substantial uncertainty that dominates the error budget. There are a few cases in which a steeper spectral index than $-$0.7 has been observed in an outburst, e.g. Swift~J1357.2$-$0933 \citep[$\alpha \sim -$1.4 in quiescent mid-IR to UV SED,][]{shahbaz}, XTE~J1550$-$564 \citep[$\alpha \sim -$1.5 in the NIR jet spectra during the transition from the soft to the hard state,][]{russell2010}, XTE~J1118+480 \citep[$\alpha = -$1.38$\pm$0.08 in the optically thin SEDs at NIR bands,][]{russell2013}, Swift~J1910.2$-$0546 \citep[$\alpha=$ -1.83$\pm$0.63 in the optical bands during reflares,][]{saikia1910apj}. A steeper spectral index is generally explained by a thermal, Maxwellian distribution of sub-relativistic electrons in a weak outflow or jet \citep{russell2010,russell2013,shahbaz,kol}, but it is unlikely at such high X-ray luminosities. In the case of IGR J17091, the power-law indices of the optical spectra, constructed with A$_V$ = 3.83 mag, are too steep (see Table \ref{table:sed}) to be consistent with the outer regions of a cold accretion disk or even a population of thermal electrons in a weak jet. 

We then reconstructed the SEDs using the most recent $N_{\mbox{\scriptsize H}}$ measurement from \citet{wang2}, which translates to A$_V$ = 5.36. Since this value was derived from NICER data, which provides better sensitivity to the soft X-ray spectrum and a more accurate estimate of $N_{\mbox{\scriptsize H}}$, we consider it to be the most reliable value of extinction currently available. In contrast, previous estimates based on RXTE data were more sensitive to harder X-ray energies, making the extinction correction less precise. The optical spectra obtained using this updated $N_{\mbox{\scriptsize H}}$ value are significantly flatter and smoother, with spectral slopes that align well with the expectations for an accretion disk ($\alpha_{V-i^{\prime}}$ ranging from $-$1.15 to $+$0.05, see Table \ref{table:sed}). Our optical analysis supports the higher extinction toward the source, with $N_{\mbox{\scriptsize H}} = (1.537 \pm 0.002) \times 10^{22} \rm cm^{-2}$ \citep{wang2} providing a significantly better fit to the optical SEDs than earlier estimates.\\

Generally, we can assume an upper limit for the slope of the optical SED ($\alpha_{V-i{^{\prime}}}$ < 2.0) owing to the Rayleigh-Jeans limit. A rough lower limit of $\alpha_{V-i{^{\prime}}} > -1$ is expected during outburst, as it roughly corresponds to the limit for the optically thin synchrotron emission using typical values of the electron particle distribution index. Sometimes, at the end of an outburst when the source is close to quiescence, the lowest plausible outer disc temperature could be as cold as $\sim$4500 K, which corresponds to $\alpha = -$ 2.4 \citep[e.g., Cen X-4 at the end of its misfired outburst in 2021,][]{cen}. In Fig. \ref{fig:dave}, we plot the optical SED slopes against different values of extinction. The three solid horizontal lines depict the expected limits on the optical spectral index $\alpha$, while the two slanted lines show the lowest and highest optical colors observed in IGR~J17091 (for the case of $A_{V}$ = 5.36, see Fig. \ref{fig:cmd}, and extrapolated to other extinction values in Fig. \ref{fig:dave}). Given the limits expected on the spectral slope, we can constrain the extinction $A_{V}$ to be in the rough range of 4.3 to 6.6. This corresponds to an $N_{\mbox{\scriptsize H}}$ range of 1.3--1.9 $\times 10^{22}\, \rm cm^{-2}$. This is much higher than the mean Galactic value, which could suggest the presence of dust clouds localized along the line of sight. The value of the extinction, as expected from the optical spectra, while somewhat above the most commonly used value of 
$N_{\mbox{\scriptsize H}}=(1.1\pm0.3)\times 10^{22}\, \rm cm^{-2}$ \citep[][etc.]{capi,rodri,krimm}, is consistent with more recent X-ray studies of the source \citep[]{xu,wang2}. 

\subsection{Distance to IGR J17091} \label{sec:distance}

\begin{figure}
    \centering
    \includegraphics[width=1.0\columnwidth, angle=0]{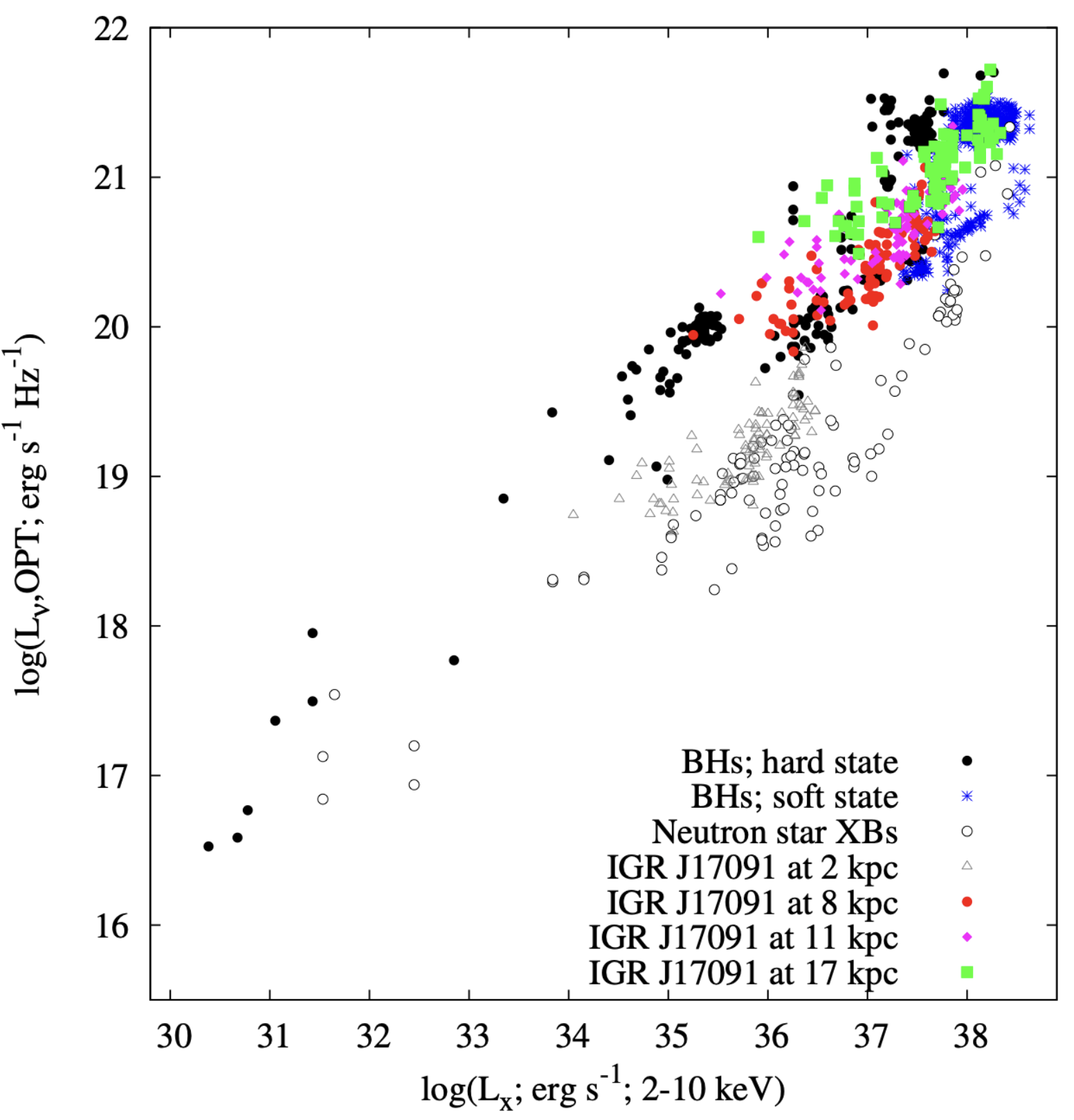}
    \caption{Global optical/X-Ray correlation of IGR~J17091 against a large sample of BH (filled black circles for hard state, blue crossed-diamonds for soft state) and NS LMXBs (black open circles), assuming the distances of 2 kpc (grey, open triangles), 8 kpc (red circles), 11 kpc (magenta diamonds) and 17 kpc (green squares).}
    \label{fig:globalox}
\end{figure}

The distance to IGR~J17091 remains uncertain, with various studies suggesting a wide range of possible values. \citet{rodri} estimated a distance of 11–17 kpc, based on X-ray spectral modeling and comparisons with other black hole X-ray binaries. \citet{diego2011} suggested a lower limit of 11 kpc, arguing that IGR~J17091 shares similarities with GRS 1915+105, which is located at ~8–10 kpc. Similarly, \citep[][]{rodri} noted that the position of IGR~J17091 in the radio/X-ray correlation plot is compatible with that of the other BHXBs only for distances greater than 11 kpc.  Meanwhile, \citet{pahari} proposed a distance of ~20 kpc, based on spectral fitting and assuming a high-inclination system. More recently, \citet{wang2} suggested a distance closer to 10 kpc, considering updated extinction estimates and X-ray luminosity constraints. Given these discrepancies, the true distance remains an open question, significantly affecting estimates of the source’s luminosity and accretion properties.\\

The global optical/X-ray correlation observed in BHXBs can provide an independent, albeit indirect, estimate of the distance to a source. This method relies on the well-established correlation between optical and X-ray luminosities in BHXBs and NSXBs during outburst, which follows a power-law relationship \cite{Russell2006,Russell2007}. By comparing the observed fluxes with the expected luminosities from known sources at well-constrained distances, one can infer a plausible distance range. This technique has been applied to several BHXBs where direct distance measurements are unavailable, offering a useful alternative to constraining their physical properties \citep[e.g.][]{Russell10,saikia1716,saikia1910mnras}.

We compare the optical/X-ray correlation of IGR~J17091 with the comprehensive sample of BHXB and NSXB in Fig. \ref{fig:globalox}, utilizing quasi-simultaneous optical and X-ray data from \cite{Russell2006,Russell2007}. Typically, NS systems are approximately 20 times optically fainter than BH systems \citep{Russell2006,Bernardini2016}.Our findings indicate that any distance within the range of 8--17kpc places the IGRJ17091 data, comprising both hard and soft state observations, well within the region occupied by other BHXBs in the global plot. For distances closer than $\sim$ 2 kpc, the data would be more consistent with the population of NSXBs. So, if the compact object in IGR~J17091 is indeed a BH, the global optical/X-ray correlation would suggest a rough minimum distance $\sim$ 8 kpc. In addition, if the distance to IGR~J17091 is as great as 17 kpc, the brightest optical data observed during the peak of the 2022 outburst would correspond solely to historical soft state data, not hard state data. As the 2022 outburst peak did not occur during a soft state \citep[as confirmed by][the 2022 outburst peak was during an intermediate state]{wang1}, and the jet is not evident in the optical SEDs or CMD at the 2022 peak, the global optical/X-ray correlation plot would argue against a large distance. In summary, our analysis is consistent with existing distance constraints available in the literature.

\subsection{Minute-scale optical variability} \label{sec:shortterm}

IGR~J17091 exhibits highly structured X-ray variability on sub-second timescales. To investigate whether such structured behavior is present in the optical, we conducted minute-timescale observations over 11 days during the 2022 outburst, which represents the best achievable time resolution with LCO given the target brightness to obtain a reasonable S/N. Across all observing nights, we find indications of structured optical variability, with fractional rms values ranging from approximately 1\% to 8\% (see Table \ref{table:rms}). However, we detect statistically significant (3$\sigma$) variability on only three of these nights (MJD 59680, 59686 and 59708). Previous studies of other BHXBs have reported similar optical fractional rms values for similar time resolutions. For many BHXBs in outburst, a 3--5 per cent rms has been observed in both the hard state \citep[e.g., Swift~J1753--0127, XTE~J1118+480, MAXI~J1535--571, GRS~1716--249;][]{gandhi2009,hynes2009,Baglio2018,saikia1716} and the soft state \citep[e.g., GX~339--4 and GRO~J1655--40;][]{hynes1998,obrien2002,cadollebel2011}. However, it should be noted that the optical fractional rms of many BHXBs, particularly in the hard and flaring states when the jet is dominant, can be as high as 5-20 per cent on shorter timescales \citep[e.g., GX~339--4, Swift~J1357.2--0933, V404~Cyg, MAXI~J1820+070;][]{gandhi2009,gandhi2010,cadollebel2011,gandhi2016,paice2019,2021MNRAS.504.3862T}.

Despite the limited detections, this remains the most detailed optical dataset available for IGR J17091 to date. While some light curves show structured, minute-timescale deviations, the cadence and signal-to-noise of the data do not allow us to construct a meaningful power spectrum, and therefore an optical QPO cannot be confirmed. We thus interpret these features conservatively as significant short-timescale variability rather than quasi-periodic behaviour. For context, optical and infrared QPOs have been reported in other BHXBs such as GX 339–4, where \citet{Kalamkar2016} identified a 0.08 Hz optical/IR QPO corresponding to half of the 0.16 Hz X-ray QPO, and multi-wavelength QPOs (optical, UV, and X-ray) have been observed in XTE J1118+480 at a common frequency \citep[][]{hynes2003}. Future higher-cadence, higher-S/N optical monitoring of IGR J17091 will be essential to determine whether similar multi-wavelength QPO behaviour is present.


Previous studies suggest that rapid multi-wavelength variability can occur at high accretion rates and luminosities, where disk instabilities drive disk-jet cycles on timescales of 10 to 1000 seconds \citep{vinnature}. Although such variability is typically observed in X-rays, under specific conditions, including system inclination, orbital parameters, obscuration, and jet brightness, these instabilities may also manifest in UV, optical, and infrared bands. In addition, they have been linked to radio-emitted ejections \citep{fen97}. However, in the case of IGR J17091, the expected luminosities and accretion rates are relatively low, making this explanation less straightforward. An alternative possibility is that these modulations represent smeared X-ray variability, reprocessed in the accretion disk and observed as structured optical fluctuations. If so, this could provide a potential link between X-ray heartbeats and the optical variability. However, confirming this scenario requires higher-time-resolution optical observations of the source.

\section{Conclusion} \label{sec:con}

This study presents the first long-term optical monitoring of the black hole X-ray binary IGR~J17091-3624, covering its outbursts in 2011, 2016, and 2022. Using data from the Las Cumbres Observatory (LCO), VISIR/VLT, Swift/XRT, RXTE, NICER, and combined with NIR monitoring from the literature, we investigate the physical processes contributing to the optical emission of this unique source.

Our analysis suggests that the optical emission from IGR~J17091 originates from an accretion disk, which could be either an X-ray-irradiated disk, a viscous disk, or a combination of both. We find that the optical flux is strongly correlated with the X-ray flux ($F_{\mbox{\scriptsize opt}} \propto F_{\rm X}^{0.40\pm0.04}$), with the power-law index indicating that the optical emission originates in the accretion disk. By comparing this correlated behavior with other BHXBs, we find that a distance of 8--17 kpc places the source within the typical range of BHXBs, consistent with previous distance estimates.

Employing the most commonly used extinction value, $N_{\mbox{\scriptsize H}} = (1.1 \pm 0.3) \times 10^{22} \, \rm cm^{-2}$, the optical spectral energy distributions appear steep and red, too extreme to be explained by standard accretion disk or jet models. However, applying the updated hydrogen column density of $N_{\mbox{\scriptsize H}} = (1.537 \pm 0.002) \times 10^{22} \, \rm cm^{-2}$ yields de-reddened spectra with more physically plausible power-law indices. From the CMD and expected optical slopes, we estimate the extinction to lie within $A_{V} = 4.3$–$6.6$ mag, corresponding to $N_{\mbox{\scriptsize H}} = (1.3$–$1.9) \times 10^{22} \, \rm cm^{-2}$. These corrections produce flatter SEDs consistent with emission from an irradiated disk, in agreement with the observed CMD trend.

Furthermore, our high-cadence optical data allowed us to investigate potential connections between optical variations and the highly structured X-ray variability observed with NICER. We find that the fractional rms deviation in the optical flux reaches up to a few per cent. We detect significant minute-timescale optical variability in the high-cadence observations, although the current data are insufficient to assess whether this variability is periodic. These variations may represent reprocessed and smeared signatures of underlying X-ray variability, hinting at a potential coupling between the X-ray and optical emission in these heart-beating sources. Confirming this scenario, however, will require higher time-resolution optical data. The observed optical variability of IGR~J17091 provides an invaluable opportunity to explore the physical properties of the accretion disk and the interplay between optical and X-ray emission in such exotic systems. This study highlights the need for even higher-cadence optical and infrared observations with larger telescopes, along with simultaneous X-ray monitoring, to capture significant optical/IR modulations and better understand their origins.

\section*{Acknowledgements}

This material is based upon work supported by Tamkeen under the NYU Abu Dhabi Research Institute grant CASS (Center for Astrophysics and Space Science). FG acknowledges support by PIP 0113 and PIBAA 1275 (CONICET). FG was also supported by grant PID2022-136828NB-C42 funded by the Spanish MCIN/AEI/ 10.13039/501100011033 and “ERDF A way of making Europe”. MCB acknowledges support from the INAF-Astrofit fellowship. This work uses data from the Faulkes Telescope Project, which is an educational partner of the Las Cumbres Observatory (LCO). The Faulkes Telescopes are maintained and operated by LCO. This work also uses data supplied by the UK \emph{Swift} Science Data Centre at the University of Leicester, and the MAXI data provided by RIKEN, JAXA and the MAXI team.

\section*{Data Availability}

The optical data from Faulkes/LCO employed in this article will be shared on reasonable request to the corresponding author. All of the \textit{NICER} and \textit{Swift} data are publicly accessible through the HEASARC portal \url{https://heasarc.gsfc.nasa.gov/db-perl/W3Browse/w3browse.pl} and the UK Swift Science Data Centre \url{https://www.Swift.ac.uk/archive/index.php}. The MAXI data employed in this article are publicly available at  \url{http://maxi.riken.jp/top/index.html}.



\bibliographystyle{mnras}
\bibliography{bib} 





\bsp	
\label{lastpage}
\end{document}